\documentclass[a4paper,11pt]{article}
\pdfoutput=1 

\usepackage{jinstpub} 
\usepackage[export]{adjustbox}

\title{\boldmath Radiation Hardness Study of the ePix100 Sensor and ASIC under Direct Illumination at the European XFEL}

\author[a,b,1]{I. Kla\v{c}kov\'a,\note{Corresponding author.}}
\author[a]{K. Ahmed}
\author[c]{G. Blaj}
\author[a]{M. Cascella}
\author[a]{V. Cerantola}
\author[c]{C. Chang}
\author[c]{A. Dragone}
\author[a]{S. G\"{o}de}
\author[a]{S. Hauf}
\author[c]{C. Kenney}
\author[c]{J. Segal}
\author[a]{M. Kuster}
\author[b]{A. \v{S}ag\'atov\'a}

\affiliation[a]{European XFEL, Holzkoppel 4, 22869 Schenefeld, Germany}
\affiliation[b]{Slovak University of Technology, Faculty of Electrical Engineering and Information Technology, Ilkovi\v{c}ova 3, 81219 Bratislava, Slovakia}
\affiliation[c]{SLAC National Accelerator Laboratory, Sand Hill Road 2575, Menlo Park, California 94025, U.S.A.}

\emailAdd{ivana.klackova@xfel.eu}

\abstract{The ePix detector family provides multiple variants of hybrid pixel detectors to support a wide range of applications at free electron laser (FEL) facilities. The ePix detectors are by design versatile and easily re-configurable camera systems with common mechanical, electrical and data acquisition interfaces. Operation of detectors at FEL sources providing high brilliance, high repetition rate and ultra-short X-ray pulses poses a high risk of radiation damage to exposed detector components, such as the sensor and the readout Application Specific Integrated Circuit (ASIC). Knowledge about radiation-induced damage is important for understanding its influence on the quality of scientific data and the lifetime of the detector.

We present the results of a systematic study of the influence of radiation induced damage on the performance and lifetime of an ePix100a detector module using a direct attenuated beam of the European X-ray Free Electron Laser Facility (European XFEL) at $9\,\text{keV}$ photon energy and an average power of $10\,\mu\text{W}$. An area of $20\,\text{pixels}\times20\,\text{pixels}$ was irradiated with an average photon flux of $\approx 7\times 10^{9}\,\text{photons}/\text{s}$ to a dose of approximately $(760{\pm 65)}\,\text{kGy}$ at the location of the $\text{Si}/\text{SiO}_2$ interfaces in the sensor.

A dose dependent increase in both offset and noise {of the ePix100a detector} have been observed originating from an increase of the sensor leakage current. Moreover, we observed an effect directly after irradiation resulting in the saturation of individual pixels by their dark current. Changes in gain are evaluated one and half hours post irradiation and suggest damage to occur also on the ASIC level. Based on the obtained results, thresholds for beam parameters are deduced and the detector lifetime is estimated with respect to the requirements to the data quality in order to satisfy the scientific standards defined by the experiments. We conclude the detector can withstand a beam with an energy up to $1\,\mu\text{J}$ at a photon energy of $9\,\text{keV}$ impacting on an area of $1\,\text{mm}^{2}$. The detector can be used without significant degradation of its performance for several years if the incident photon beam intensities do not {exceed} the detector’s dynamic range by at least three orders of magnitude. Our results provide valuable input for the operation of the ePix100a detector at FEL facilities and for the design of future detector technology.}


\arxivnumber{xxxx.yyyyy} 

\begin{document}
\maketitle
\flushbottom

\section{Introduction}
\label{sec:intro}

To fully exploit the research and scientific possibilities of the European XFEL \cite{Decking:2020a}, novel detectors have been developed in order to satisfy the challenging FEL requirements, e.g. single photon sensitivity at the specified energy range, time resolution at the level of individual FEL pulses or a high dynamic range~\cite{Graafsma:2009a, Kuster:2014a}. The European XFEL is a high repetition rate facility, delivering pulse trains with a repetition rate of $10\,\text{Hz}$. Every train contains up to $2700$ high intensity X-ray pulses separated by $220\,\text{ns}$ each. An X-ray beam with such characteristics poses a high risk of damaging the detector systems by radiation, thus creates the need for radiation hard detectors. For a detector operated in a scattering geometry close to the sample and being directly illuminated by scattered X-rays, the absorbed doses can amount in up to $1\,\text{GGy}$ using a silicon sensor of $500\,\mu\text{m}$ thickness when considering three years of facility operation \cite{Graafsma:2009a}. While some of the European XFEL detectors have been optimized to incorporate a higher level of radiation hardness into their design, e.g. the Adaptive Gain Integrating Detector (AGIPD)~\cite{Zhang:2014a} and the Large Pixel Detector (LPD)~\cite{Koch:2013a}, other detectors, not specifically built for operation at the European XFEL, have been tested for and offer a certain level of radiation tolerance, e.g. the JUNGFRAU~\cite{Jungmann:2015a} detector. The ePix100a detector, designed for applications at FEL facilities \cite{Blaj:2015a} was studied to evaluate its radiation hardness at the European XFEL. The aim of the presented study is to determine the level of damage caused by the X-ray laser beam, understand and characterize the radiation-induced damage effects and validate their impact on the quality of scientific data.

\section{Radiation-Induced Damage Effects on Silicon Detectors}
\label{sec:RadEff-theory}
Operating silicon sensors in the harsh radiation environment of FELs can have severe implications on a detectors' performance and its life time. Due to their high peak brilliance, FELs can deliver up to $10^{12}\,\text{photons/pulse}$ to the sample interaction region. Due to the direct X-ray illumination of the sensor, radiation tolerance of the sensor and Application Specific Integrated Circuit (ASIC) are of high importance. Sensors based on metal-oxide-semiconductor (MOS) structures, where charge flow takes place close to the surface, are known to be {especially} sensitive to damage induced by ionizing radiation. Therefore, thorough studies of the influence of the FEL radiation on silicon sensors have been conducted during the design and development phase of the first generation MHz detectors for the European XFEL. Results reported by Zhang et al.~\cite{Zhang:2011a, Zhang:2012a, Zhang:2014a} give insight on the parameters determining the damage depending on the delivered dose and their influence on the operation of different silicon sensor designs. These studies provided valuable input for detector designs minimizing radiation damage effects and optimization of sensor operation parameters as reported by Schwandt et al.~\cite{Schwandt:2012a, Schwandt:2013a}. Klanner et al.~\cite{Klanner:2013a} provide an exemplary overview of the AGIPD detector~\cite{Allahgholi:2019a} related sensor design challenges. The studies cited above provide useful knowledge and valuable observations for the here presented experiment results. 

In general two fundamentally different damage mechanisms dominate the radiation/particle interaction with silicon. These are displacement damage (bulk damage) and surface damage. Bulk damage results in point defects (Frenkel-pairs) or agglomeration of defects caused by {highly} energetic hadrons, leptons or higher energetic gamma rays knocking out the primary atom from its lattice position (interstitial) through a non-ionizing (NIEL) interaction~\cite{Affolder:2011a,Lindstroem:2003a}. { The displacement energy $E_\text{d}$ required to create such a defect is in the range between $10\,\text{eV}$ and $36\,\text{eV}$ \cite{Bourgoin:1976a,Holmstroem:2008a}. $E_\text{d}$ depends amongst other parameters on the lattice orientation. Since the energy transfer of a photon with an energy less than $20\,\text{keV}$ to the silicon atom is significantly below $E_d$, displacement damage is negligible at these photon energies. Consequently, surface damage is the dominating damage mechanism of relevance for our study.}

In contrast to bulk damage, surface damage originates in ionization energy losses of X-ray photons or charged particles and subsequently leads to an accumulation of space charges in or close to an interface between e.g. an insulating or dielectric layer and silicon~\cite{Ling:99a}. A typical example are interfaces between $\text{SiO}_2$ and silicon when $\text{SiO}_2$ is used as gate oxide or field oxide as insulating layer between semiconductor structures. The density of the created space charge is proportional to the amount of energy absorbed at or close to the interface. The accumulation of space charge in turn can significantly influences the performance properties of the sensor, like leakage current, noise and dynamic range. For an in-depth overview about surface damage mechanisms, we refer the reader to the available literature \cite{McLean:1987a, Dressendorfer:1989a, oldham1999ionizing} and references therein.


\begin{figure}
\centering
\includegraphics[width=0.35\textwidth]{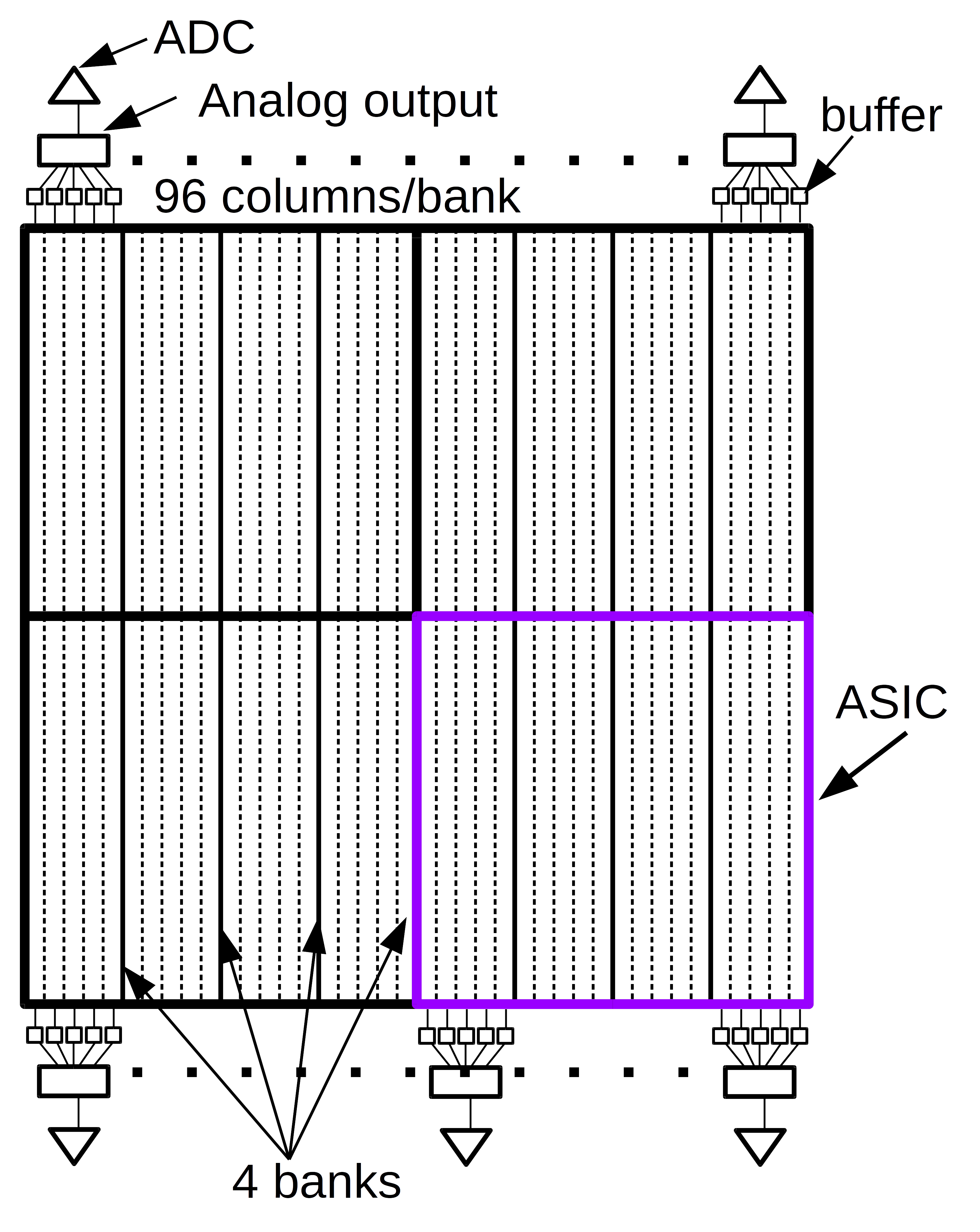}
\qquad
\includegraphics[width=0.5\textwidth]{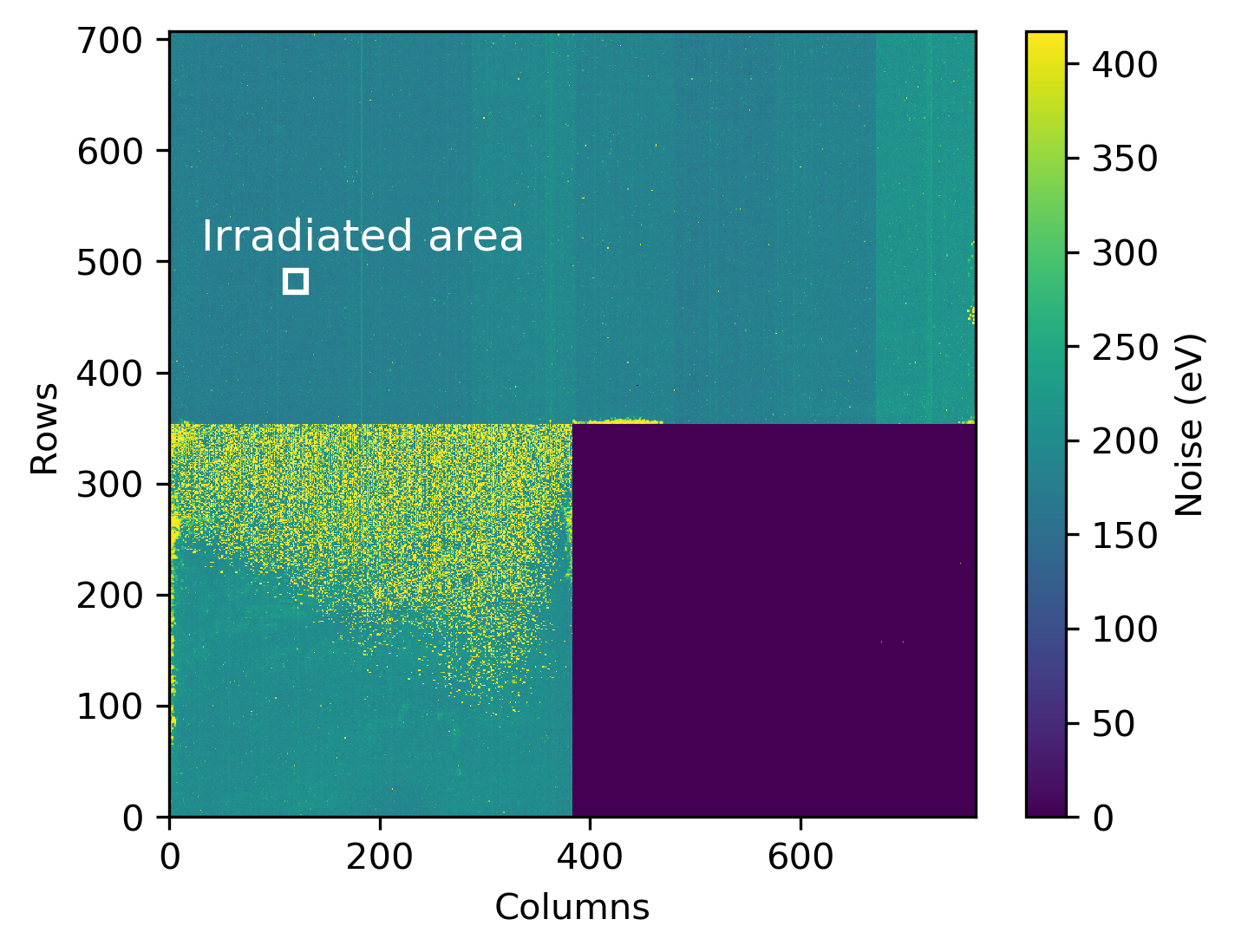}
\caption{\label{fig:ASIC_beamSpot} Left: Schematic view of the ePix100a sensor as seen from the experiment's interaction region, showing the location of the signal readout nodes, the arrangement of the 4 ASICs required to read out one sensor module with a size of $352\,\text{pixels} \times 384\,\text{pixels}$. Each ASIC is divided into four banks accommodating 96 columns multiplexed to a single analog output digitized by an external ADC. {Four sensor modules and their ASICs are arranged in a rectangular geometry to build one complete detector module consisting of $704\,\text{pixels} \times 768\,\text{pixels}$.} Right: Noise map of the ePix100a visualizing  placement and size of the area irradiated by the FEL beam ({white} square). The area on the bottom left is showing a higher noise level in comparison to the two modules on the top. The bottom right module is not providing data. The image qualitatively illustrates the performance before irradiation.}
\end{figure}

\section{The ePix100a Detector Module}
\label{subsec:ePix}
The ePix100a is a backside illuminated direct detection hybrid pixel detector optimized for low noise applications in the energy range between $2\,\text{keV}$ and $18\,\text{keV}$ \cite{Blaj:2016a}. The ePix detector family is based on a modular design, where each ePix100a detector module consists of a fully depleted silicon sensor, which is flip-chip bonded to 4 ePix ASICs, and connected to front-end electronics, the cooling system, its mechanics and housing.

\begin{figure}
\centering
  \includegraphics[width=0.53\textwidth]{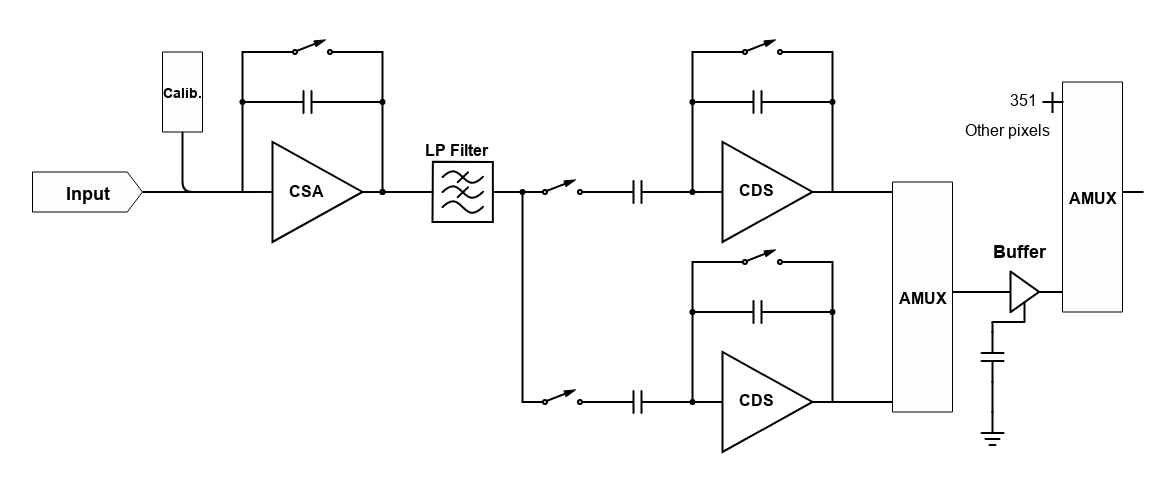}
    \hfill
    \includegraphics[width=0.43\textwidth]{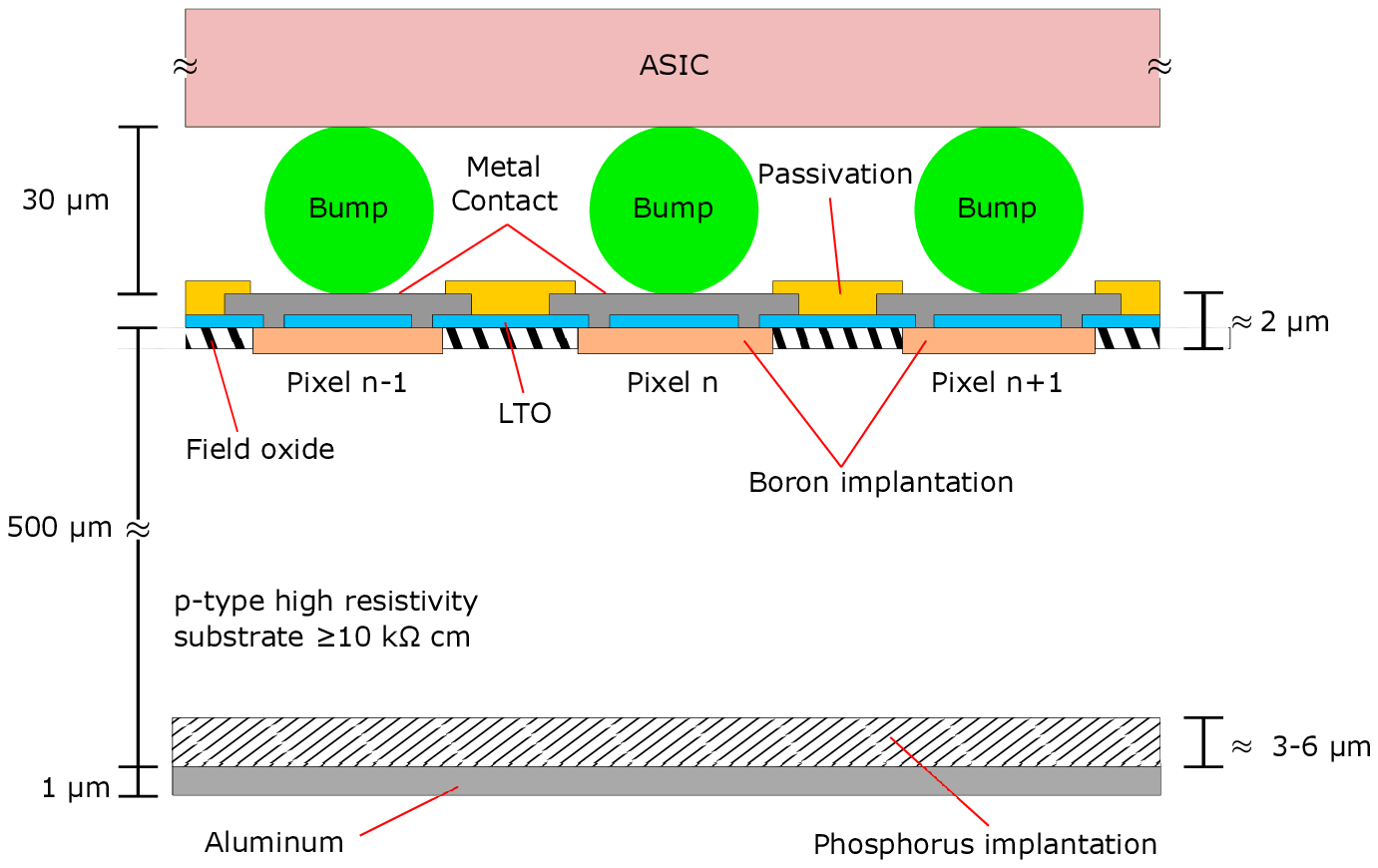}
\caption{\label{fig:SensorCrosssection}{Left: Schematic drawing of the ePix100a ASIC pixel layout. Right: }Schematic view of the ePix100a sensor cross-section including interconnection bump bonds and the readout ASIC. The sensor is illuminated from the backside, i.e. from the bottom. Please note that this drawing is not to scale.}
\end{figure}

One single ASIC provides the readout and signal processing architecture for a sensor region of $352\,\text{pixels} \times 384\,\text{pixels}$, each of a size of $50\,\mu\text{m} \times 50\,\mu\text{m}$. The analog signal provided by a sensor pixel is processed by a low noise charge integrator and subsequently low pass filtered before the signal passes a correlated double sampling (CDS) stage for baseline correction and noise filtering and is finally stored in a buffer. Further processing of the analog signal is organised in a column parallel fashion. The analog output of the pixels of one bank accommodating $96$ columns is multiplexed to a single analog output node, before it is digitized by an external {$14\,\text{bit}$} sigma-delta analog to digital converter (ADC). { An overview of the ePix100a ASIC pixel layout is schematically shown in the left part of figure~\ref{fig:SensorCrosssection}}. The dynamic range of the ePix100a detector, defined by the number of ADC digitisation levels available for photon detection, allows to measure up to $220\,\text{ke}^-\approx100\times 8\,\text{keV}$ photons per pixel at a maximum frame rate of $240\,\text{Hz}$. Figure~\ref{fig:ASIC_beamSpot} (left panel) illustrates the geometric arrangement of the readout banks and the ADCs. The analog output nodes are arranged on the top and bottom sides of the ASICs and sensor. For a detailed description of the ePix ASIC, detector design and a performance review, we refer the interested reader to Markovic et al.~\cite{Markovic:2014a}, Blaj et al.~\cite{Blaj:2016a} and Nishimura et al.~\cite{nishimura2016design}. 

Figure~\ref{fig:SensorCrosssection} shows a schematic view of the vertical cross-section of the ePix100a sensor {produced by SINTEF}. The photon entrance window is coated with aluminum acting as a light blocking filter for optical, UV and IR light. The bulk of the sensor is made of p-doped high resistivity {$<100>$} silicon. 
Due to its backside illuminated design, structures potentially vulnerable to radiation damage, i.e. interfaces to the Low Temperature Oxide (LTO) and Field Oxide, are located on the front side of the chip. The interconnection between the sensor and ASIC is provided through $30\,\mu\text{m}$ large solder bump bonds {attached to Ti/Cu metal gates. The pixels and guard rings are biased through the ASIC.}
The $500\,\mu\text{m}$ thick sensor enables the detection of X-ray photons with energies between $3\,\text{keV}$ and $13\,\text{keV}$ with a quantum efficiency $\ge 80\,\%$ and at the same time efficiently shields the underlying ASIC from X-ray radiation at low photon energies. The ePix100a camera is currently being used at the High Energy Density (HED) instrument \cite{HED:2021a} and Material Imaging and Dynamics (MID) \cite{MID:2021a,Madsen:2021b} instrument at the European XFEL. 

\section{Experiment Setup and Methodology}
\label{sec:Experiment Setup and Methodology}
The radiation damage study was performed at the HED instrument at the European XFEL. The HED instrument can provide a beam of photon pulses with an energy between $5\,\text{keV}$ and $20\,\text{keV}$ and a maximum pulse energy of approximately $1\,\text{mJ}$ \cite{HED:2021a}. For our study we used an ePix100a test module equipped with four ASICs and arranged as shown in figure~\ref{fig:ASIC_beamSpot}. Two of the ASICs were fully functional (area on the top left and right shown in the right image of figure~\ref{fig:ASIC_beamSpot}), one had a significantly higher mean noise before irradiation (image area on the bottom left part of the same image) and one was unresponsive (bottom right part of the image). The beam spot, having an area of approximately $1\,\text{mm}^2$ and covering approximately $20\,\text{pixels}\times20\,\text{pixels}$ of the ePix100a sensor, was {used to irradiate} the area with the lowest pre-irradiation noise referred to as region of interest (ROI) during the remainder of this publication (see figure~\ref{fig:ASIC_beamSpot}, right panel). 

{Exposing the detector to the direct beam could cause instantaneous and permanent physical damage to the sensor, e.g. by ablation or melting of the sensor material. As shown by Koyama et al.~\cite{Koyama:2013}, the silicon ablation threshold for a $10\,\text{keV}$ photon is $E_{th}=0.78\,\mu\text{J}/\mu\text{m}^2$. To avoid this kind of damage, we initially attenuated the FEL beam to energies well below $E_{th}$ with a configurable stack of Chemical Vapor Deposition (CVD) diamond and Si foils of various thicknesses available at the instrument. Apart from physical damage to the sensor surface, components of the pixel ASIC could be damaged by irradiation with excessive beam energies, as the ASIC [23, 39] does not implement protection circuitry for signals significantly above the detector’s dynamic range. To find a beam energy, which on the one hand is sufficient to achieve the anticipated dose rate of $200\,\text{kGy/h}$ and at the same time allows safe operation of the detector, we step by step gradually reduced the absorption of the stack until we found a configuration which allowed us to illuminate the detector up to an exposure time of $> 5\,\text{min}$ without observing failure of the irradiated pixels. By following this procedure, we found that the ePix100a pixels can withstand beam energies of up to $1\,\mu\text{J}$ for longer time periods without individual pixels experiencing permanent failure. Hence, we irradiated the sensor with a beam energy of $1\,\mu\text{J}$. As discussed later in the text, beam energy of $1\,\mu\text{J}$ causes pixels to become unresponsive during the duration of irradiation, but it does not cause permanent failure of the irradiated pixels. This beam energy setting was used deliberately to trigger pixels unresponsiveness to allow for beam position tracking (explained in section~\ref{sec:DoseCalib}).}

\begin{figure}
\centering
\includegraphics[width=0.69\textwidth]{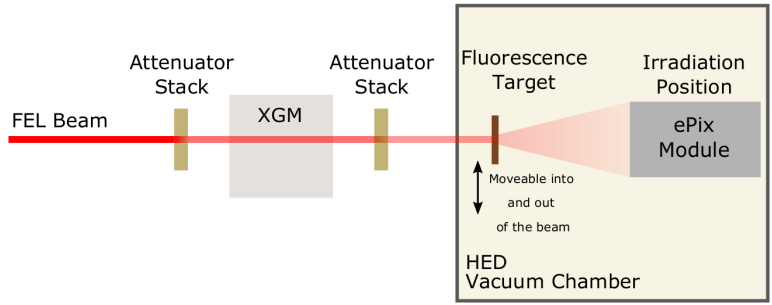}
\caption{\label{fig:EXPalignment} Experimental setup used to irradiate and calibrate the ePix100a detector module at the HED instrument (not to scale). The image shows the setup used for flat-field performance and calibration measurements {(flat-field configuration)}. The attenuated beam is directed onto a {copper} target and the ePix100a module is shifted off the beam axis. The resulting fluorescence radiation illuminates the ePix100a module homogeneously. Since we used the {copper} target in a transmission geometry, the attenuated direct FEL beam is visible on the detector in addition to the {copper} fluorescence's photons (see figure~\ref{fig:Cu_calibSpec}). During irradiation measurements the ePix100a is facing the beam and the fluorescent target is moved out of the field of view of the ePix100a module {(irradiation configuration)}. In both configurations the primary beam intensity is attenuated by two absorbing and configurable stacks located before and after the XGM.}
\end{figure}

 The beam energy was continuously monitored with the X-ray Gas Monitors (XGM) installed at the HED instrument throughout our study {(see \cite{Maltezopoulos:2019a,Sorokin:2019a} for a technical description of and details about the performance of the European XFEL photon diagnostic XGMs)}. The XGMs are capable of non-invasively measuring the single X-ray pulse energy with an absolute accuracy of $7\% - 10\%$ depending on the measured signal strength and to provide a beam position measurement with a precision between $10\,\mu\text{m}$ and $50\,\mu\text{m}$ if the beam position stays within $\pm 6\,\text{mm}$ of the absolutely calibrated reference position. In addition to the diagnostic information provided by the XGMs we used the ePix100a detector to measure the X-ray spot position and potential drifts of the X-ray spot. A schematic view of the experiment setup is shown in { f}igure~\ref{fig:EXPalignment}. 

To monitor the system noise, leakage current, gain and energy resolution of the detector, we acquired pre- and post-irradiation calibration data with $t_\text{Int}=50\,\mu\text{s}$ consisting of { dark} and flat-field measurements. {The flat-field measurements were performed by homogeneously illuminating the ePix100a} with {$\text{Cu-K}_{\alpha}$} fluorescence photons originating in a $50\,\mu\text{m}$ thick copper foil which could be moved into the FEL beam. {To acquire dark data the FEL beam was blocked and the detector was located in flat-field configuration in the dark HED vacuum chamber (see figure~\ref{fig:EXPalignment}), excluding illumination with visible light.} A summary of the FEL beam and detector parameters used during our experiment is provided in Table~\ref{tab:parameters}. Operational constraints prevented us from taking calibration data more frequently, e.g. after the completion of each individual irradiation cycle. Noise and offset were monitored on an hourly time scale shorty after the last irradiation cycle and later on, on time scales of days after the self-annealing effects slowly started to reduce in significance.

We irradiated the ePix100a module with an attenuated beam of $9\,\text{keV}$ photons. The facility was set up to deliver $100$ pulses per pulse train (i.e. $100$ pulses per $100\,\text{ms}$). We operated the detector with a frame rate of $10\,\text{Hz}$, corresponding to the typical use case of the ePix100a detector at the European XFEL. The irradiation was performed in cycles of $20\,\text{min}$ long individual exposures. Radiation induced changes of the noise and offset were monitored regularly between two consecutive cycles with two different integration time settings, $t_\text{Int}=50\,\mu\text{s}$ and $t_\text{Int}=800\,\mu\text{s}$. The choice of the $800\,\mu\text{s}$ long integration time is motivated by the higher sensitivity to potentially very small changes of the leakage current, which would manifest through a proportional change of the offset. On the other hand, $t_\text{Int}=50\,\mu\text{s}$ is a typical value used during scientific experiments at HED. In total we executed $14$ such irradiation cycles in the course of our study. During each cycle we achieved a typical dose rate of $({204 \pm 18})\,\text{kGy}/\text{h}$ translated to the depth of the $\text{Si}/\text{SiO}_2$ interfaces in the sensor. {We determined the dose rate with a Monte Carlo simulation using the measured average photon flux at the ePix100a sensor surface and the sensor geometry as input parameters, as discussed extensively in section~\ref{sec:DoseCalib}}.

Throughout our irradiation cycles the detector was operated under conditions mimicking the typical detectors' experimental usage scenario, that is at a pressure of $1\times10^{-5}\,\text{mbar}$, cooled to $-9\,^{\circ}\,\text{C}$ and biased with a voltage of $200\,\text{V}$. After the $2\,\text{days}$ long irradiation experiment the detector was stored at room temperature under ambient atmospheric conditions, and only powered and {cooled again to the same conditions} during post-irradiation follow-up performance measurements.

\begin{table}
\centering
\caption{\label{tab:parameters}Summary of the detector operation and beam line parameters as used during our irradiation experiment.}
\smallskip
\begin{tabular}{ll}
\hline
Beam parameters                        & \\ \hline
Average beam energy at the detector    & $10\,$nJ/Pulse                  \\
Photon energy                          & $9\,$keV                        \\
Number of X-ray pulses per pulse train & $100\,$Pulses             \\
Dose rate at the $\text{Si/SiO}_2$ interfaces & $200\,$kGy/h           \\
Beam intensity monitoring              & XGMs at HED beamline             \\\hline
Detector parameters                    & \\\hline 
Pixel size                             & $50\,\mu\text{m} \times  50\,\mu\text{m}$\\
Sensor size                            & $704\,\text{pixels} \times 768\,\text{pixels}$  \\
Sensor thickness (depleted bulk)       & $500\,\mu\text{m}$              \\
Irradiated sensor area                 & $20\,\text{pixels}\times20$ pixels ($1\,\text{mm}^2$)\\
Full well capacity                     & $220\,\text{ke}^-$              \\
Frame rate                             & $10\,\text{Hz}$                 \\
Integration time                       & $50\,\mu$s and $800\,\mu$s      \\ 
Bias voltage                           & $200\,$V                        \\ 
Sensor temperature                     & $-9^{\circ}\,\text{C}$          \\ 
{Operating pressure}                   & ${\leq 1\times 10^{-5}\,\text{mbar}}$ \\
\end{tabular}
\end{table}

\subsection{Detector Performance Characterization}
\begin{figure}
\includegraphics[width=0.48\columnwidth]{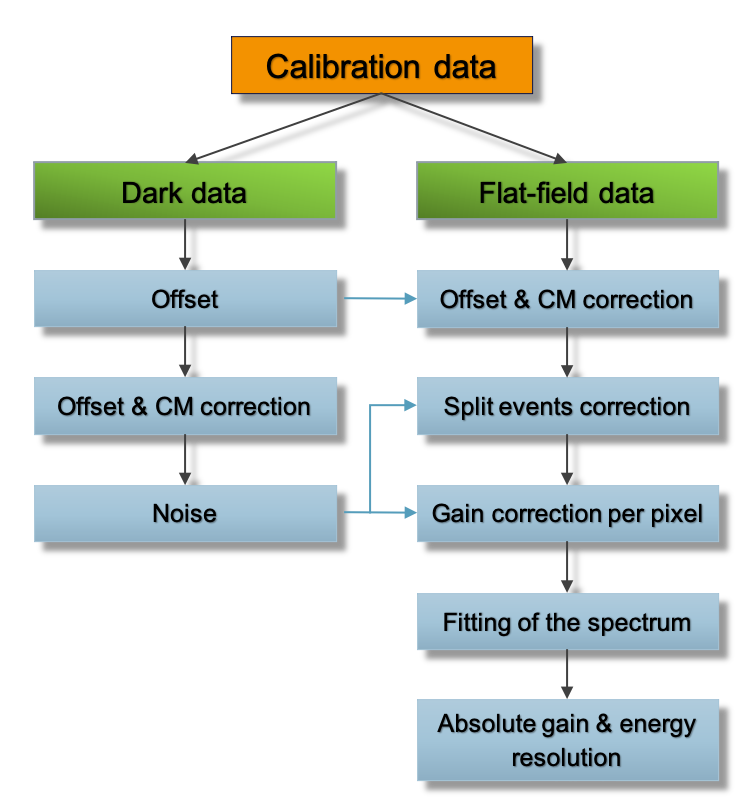}
\hfill
\includegraphics[width=0.48\textwidth]{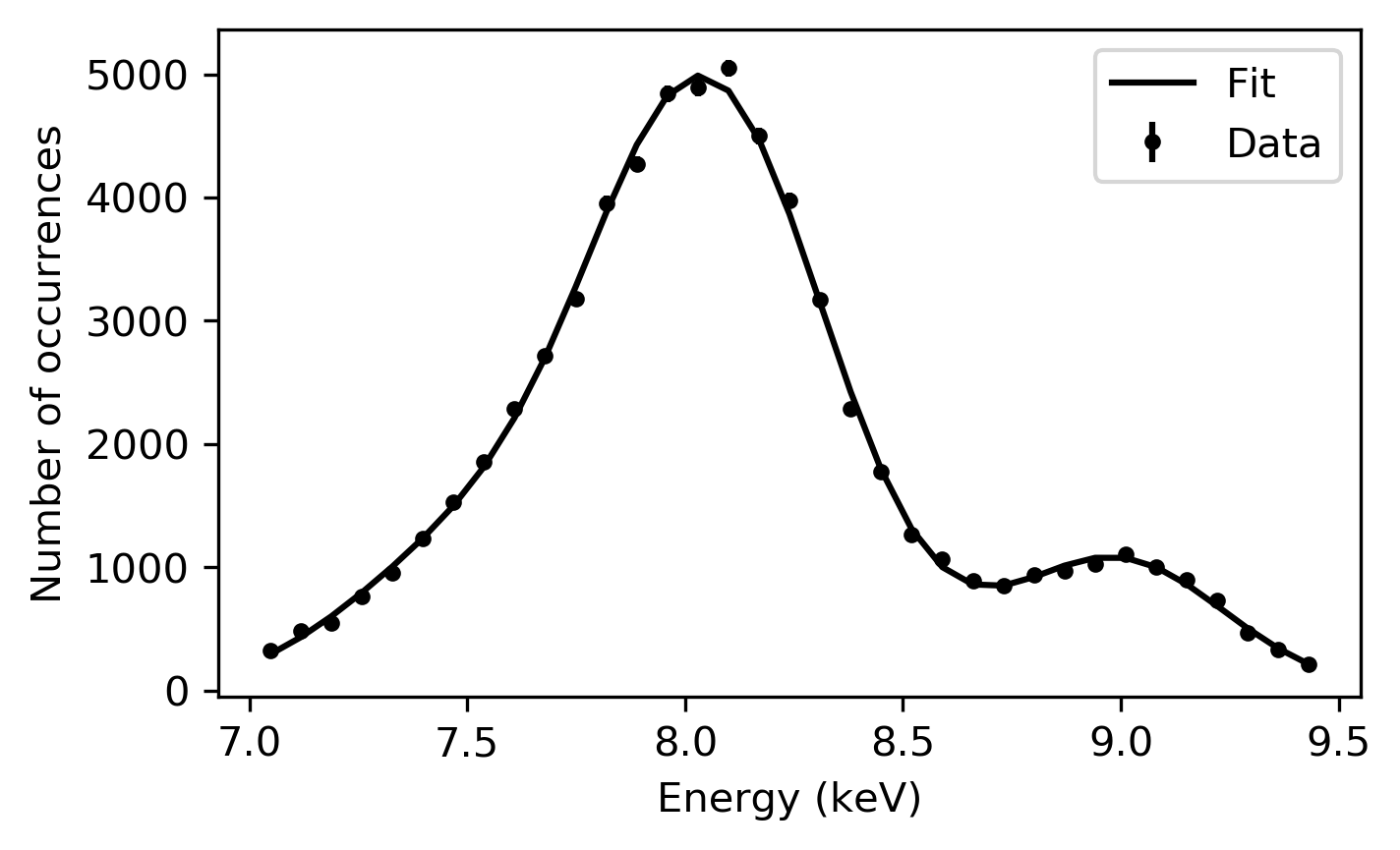}
\caption{\label{fig:FlowchartCalib}\label{fig:Cu_calibSpec}
 Left: Flowchart illustrating the data correction steps applied to the dark and flat-field data and their sequence. Results derived from the analysis of the dark data are used as input for processing the flat-field data as illustrated by blue lines. Right: Energy spectrum derived from flat-field data, showing the {$\text{Cu-K}_{\alpha}$} fluorescence line blend at {$8.041\,\text{keV}$}, a line originating in photons of the FEL beam with an energy of $9\,\text{keV}$ and a fit of {three} Gaussian peaks to the spectral data.}
\end{figure}
We characterized the offset (mean dark signal), the root mean square (RMS) noise, the pixel-to-pixel variation of the signal amplification, the pixel averaged energy resolution and the absolute gain, i.e. the {Analog Digital Unit (ADU)} to energy conversion before and after irradiation. Particular attention was paid to be able to detect offset and noise changes with a sensitivity of {$2\%$}. 

{The detector offset is commonly defined as the average value of the dark signal of the detector. The detector noise, quantifying the variations of the dark signal relative to its mean, is calculated as the standard deviation of the measured dark signal. Variations in the voltage supplying the readout electronics can cause additional offset variations affecting groups of channels with a similar amplitude, this effect is known as common mode. For a detailed description of the nature of the components contributing to the dark signal and detector noise we refer the reader to e.g. Knoll~\cite{Knoll:2011} and Lutz~\cite{Lutz:2007}.}

Flat-field data {resulted from homogeneous irradiation of the ePix100a sensor area} with {copper} fluorescence photons with an energy of $E_{\text{K}\alpha_1} = 8047.78\,\text{eV}$ and $E_{\text{K}\alpha_2} = 8027.83\,\text{eV}$, originating from a  {copper} target installed in the FEL beam  (see figure~\ref{fig:EXPalignment}). Since the {two $\text{Cu-K}_{\alpha_1}$ and $\text{Cu-K}_{\alpha 2}$ } fluorescent lines cannot be resolved with the ePix100a detector, we observed instead a blend of lines  with a {yield weighted average energy of $8041.13\,\text{eV}$}{, to which we refer to as $\text{Cu-K}_{\alpha}$ lines in the remainder of this paper.} 

The procedure used for dark and flat-field data processing is outlined in figure~\ref{fig:FlowchartCalib}. {As the first step the per pixel offset $O_{x,y}$ is calculated as mean signal for each pixel individually using $2000$ images of the dark data. With an ASIC-, column- and row-wise common mode (CM) correction we removed signal baseline variations apparent in the dark data before calculating the RMS noise for each pixel. The common mode value is evaluated from the sorted vector $S_{i}$ as
\begin{equation}
	\bar{m}_k = \left\{
	\begin{array}{ll}
	 	S_{(N-1)/2} & \mathrm{for\, an \,odd\,number\, of\, pixels} \, N\\
 		\frac{1}{2} (S_{N/2}+S_{N/2+1}) & \mathrm{for\, an \,even\,number\, of\, pixels}\, N,
	\end{array}
	\right.
\end{equation}
where $S_{i}$ is the signal measured in the pixel with the number $i$ after offset subtraction, belonging to a specific ASIC area, one line or to one column numbered with $k$. This common mode value is subsequently subtracted from the signal of the pixels belonging to the area the common mode value was calculated from. Finally the RMS noise results from}
\begin{equation}
    \sigma_{x,y}=\sqrt{\frac{1}{N}\sum_{n=0}^{N-1}(c_{x,y,n}-\overline{c}_{x,y})^{2}},
\end{equation}
{where $c_{x,y,n}$ is the offset and common mode subtracted dark signal measured in the pixels with column and line coordinates $x$ and $y$ of the image numbered $n$ . The per-pixel average of the dark signal corrected for offset and common mode is denoted with $\overline{c}_{x,y}$}. After applying the offset and a common mode correction to the flat-field data, we classified clustered signals of discrete photon events by the number of adjacent pixels where the charge is detected (split event/charge sharing correction, see~\cite{Kuster:2021a, pyDetLib:2020a} for a more detailed description of the algorithm). To achieve the best possible energy resolution for subsequent spectral analysis, clusters of events with a {pixel event} multiplicity larger than one were rejected. A per-pixel gain correction removes small pixel-to-pixel variation of the characteristics of the pre-amplifiers implemented in each pixel. The data used for calculating the energy spectrum shown in figure~\ref{fig:Cu_calibSpec} was treated in this way. 

Fitting a model consisting of three Gaussian lines (two lines for modelling the two {line} peaks and one for the low energy shoulder of the {$\text{Cu-K}_{\alpha}$ line} peak) to the spectrum derived from the data of the ROI before irradiation yields the ADU to energy conversion factor (absolute gain) $g=({70.2}\pm0.5)\,\text{eV}/\text{ADU}$ and an energy resolution of $(641\pm 45)\,\text{eV}$ full width half maximum (FWHM). The spatial distribution of  the noise and offset is consistent with a uniform distribution across the pixels inside the ROI, with a mean offset and RMS noise of $1756\,\text{ADU}$ and ${139}\,\text{eV}$, the latter value corresponds to an equivalent noise charge (ENC) of $38\,\text{e}^-$. 

\subsection{Dose Calibration}\label{sec:DoseCalib}
As described in section~\ref{sec:Experiment Setup and Methodology}, we monitored the photon flux throughout our irradiation experiment with the XGM with an estimated uncertainty {of $7\,\%$}. However, the XGM does not provide spatial information, i.e the intensity distribution of the beam profile with a position resolution equivalent or better than the spatial resolution of the ePix100a sensor. Instead we determined the beam profile from direct beam measurements acquired at low X-ray intensities {with the ePix100a}. These measurements provided spatially resolved information about the beam shape and the position of the beam within the ROI.
\begin{figure}
\centering 
\includegraphics[width=.5\textwidth]{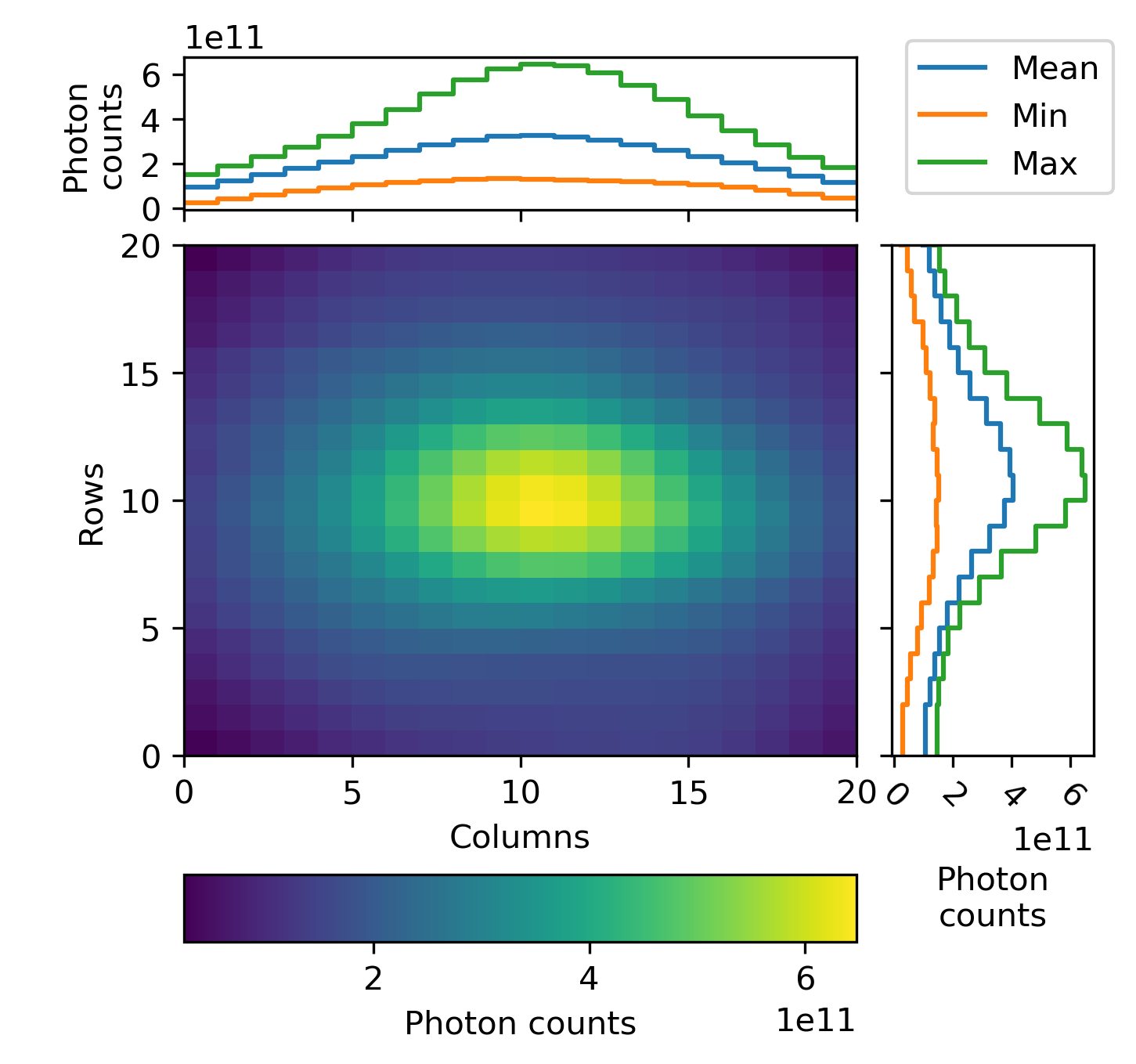}
\hfill
\includegraphics[width=.48\textwidth]{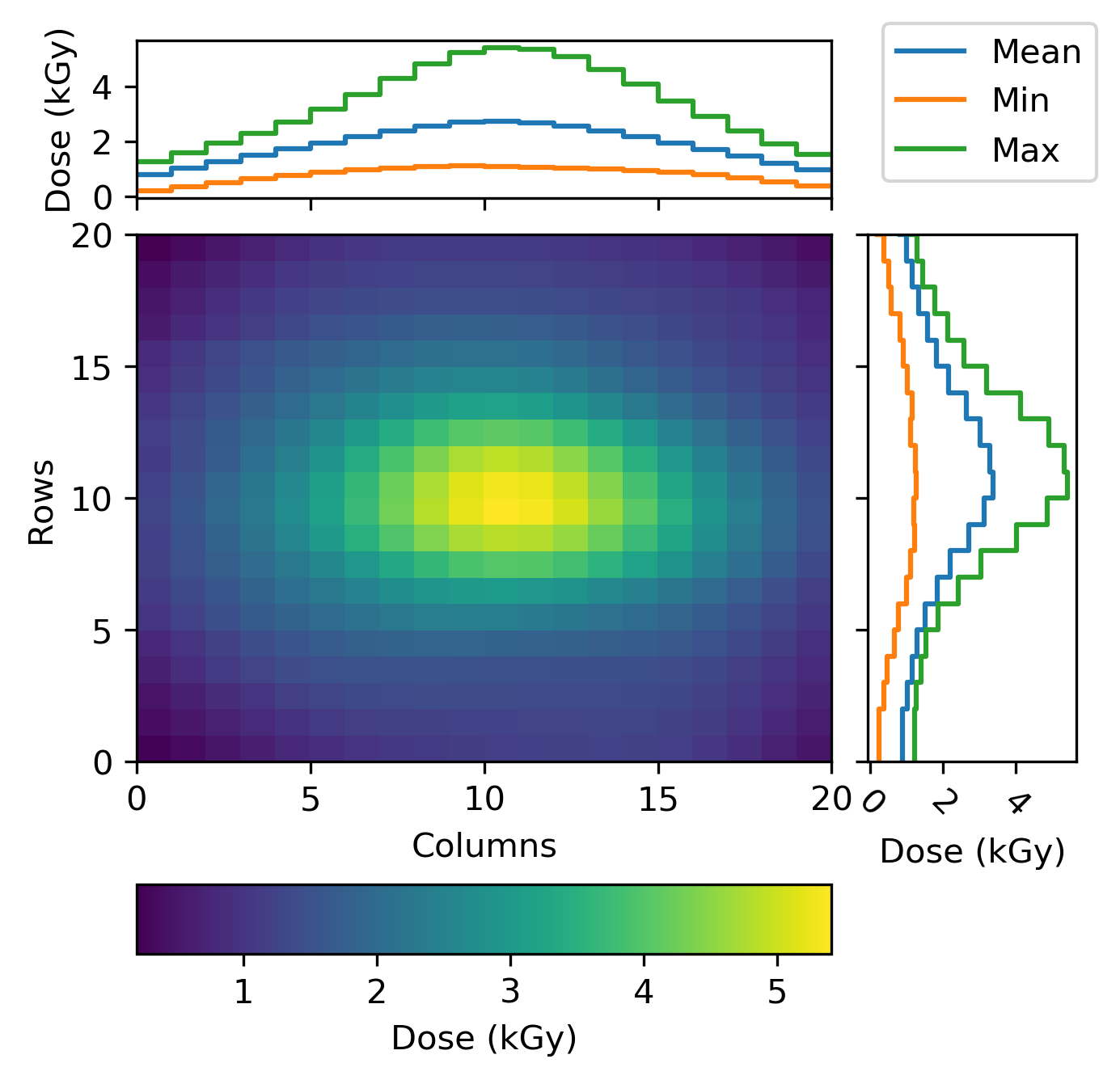}

\includegraphics[width=.49\textwidth]{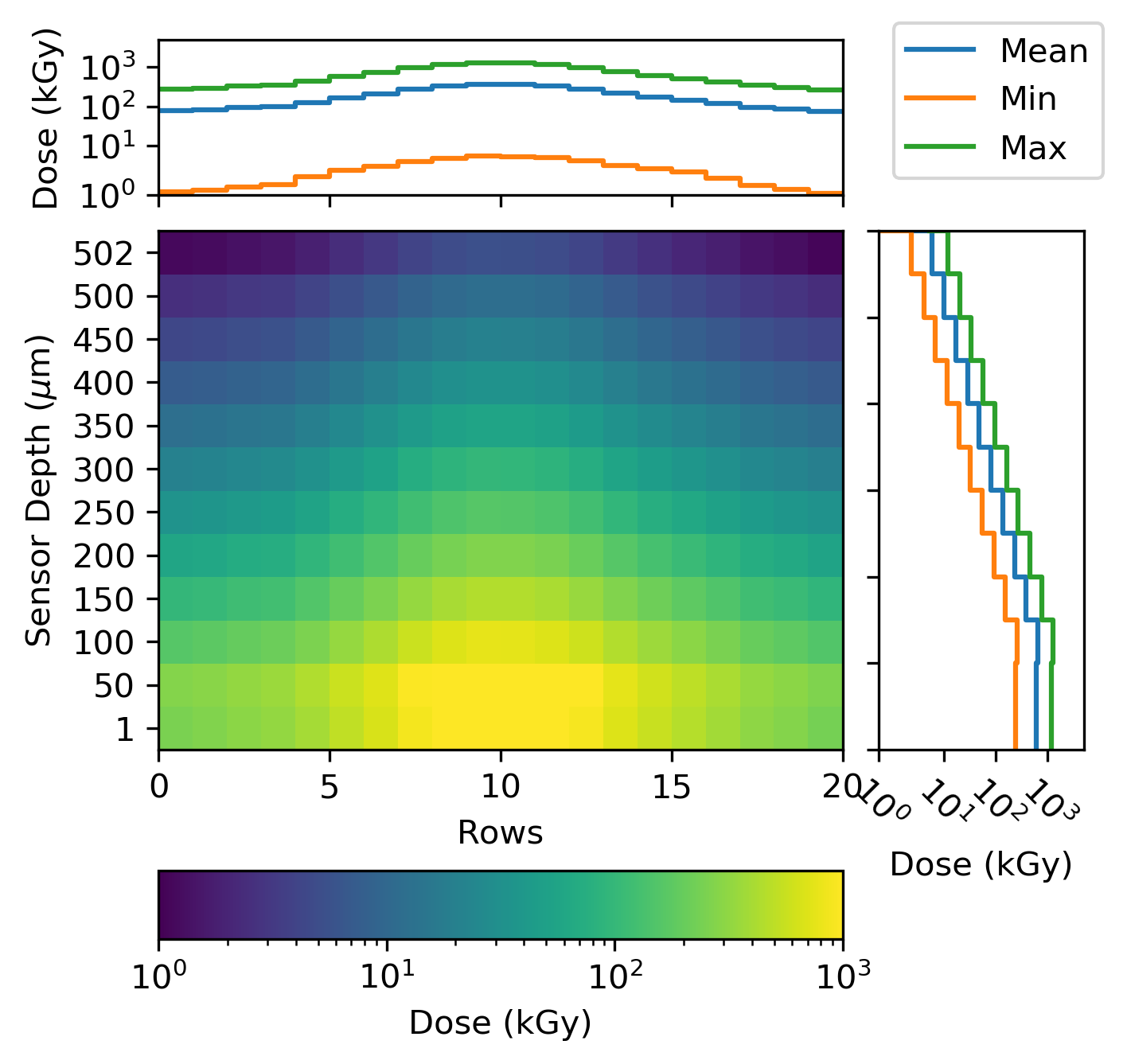}
\hfill
\includegraphics[width=.49\textwidth]{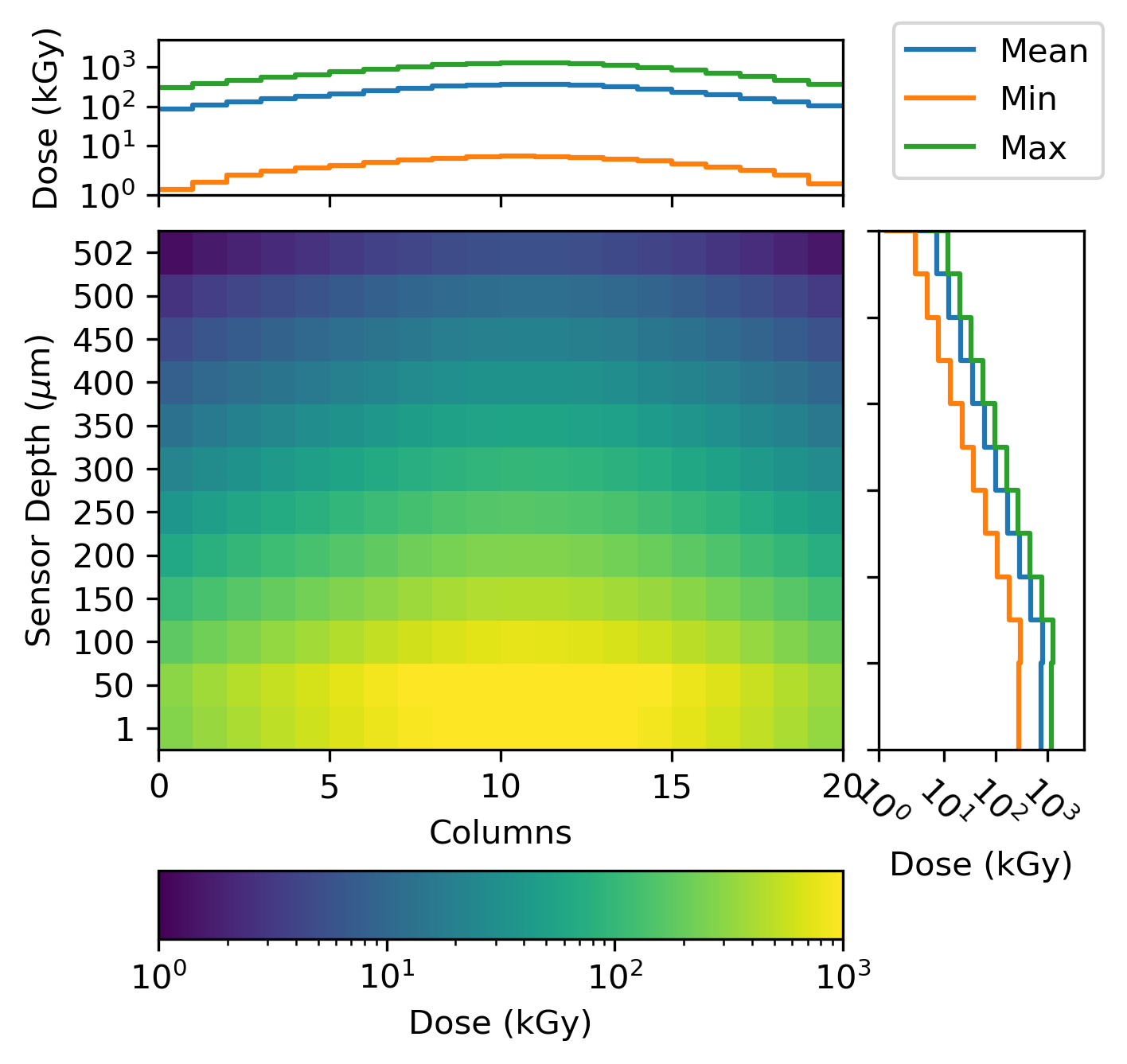}
\caption{\label{fig:dosedepth} {The top left plot shows the spatial distribution of the total number of photons in the region of interest integrated over the time of irradiation with the beam. The spatial distribution of the dose deposited at the depth of the $\text{SiO}_2$ layers is shown in the top right plot for the same region. The dose was simulated using the Monte Carlo simulation tool MULASSIS. The data shown on the top left side of this figure was used as input for the simulation.} The bottom left corner of this picture corresponds to the bottom left corner of the region of interest marked as blue rectangle and being labelled "Irradiated Area" in figure~\ref{fig:ASIC_beamSpot}. Bottom: Depth profiles of the dose along column 10 (left) and row 9 (right) of the irradiated area. The orientation of the image is the same as the schematic view of the sensor shown in figure~\ref{fig:SensorCrosssection}, i.e. the $1\mu\text{m}$ thick Aluminum entrance window is located at the bottom and the pixel structure on the top of the images. Please note that the dose values are shown on a linear scale in the top image and on a logarithmic scale in the bottom images. {The panels on top and bottom of all images show minimum (orange), mean (blue) and maximum (green) values calculated along rows, columns or the sensor depth, respectively.}}
\end{figure}

Irradiating the detector with high intensity caused {the directly irradiated pixel area} to become unresponsive for the duration of the {irradiation. T}he affected pixels provided a signal {even significantly below the detected dark signal. The average value of signal provided was $130\,\text{ADU}$.} This effect is discussed in more detail in section~\ref{sec:Experiment Results}. The position of the affected area on the sensor relative to the centre of the beam does not change over time. The importance of this effect lies in the possibility to track the position of the beam on the ePix100a sensor, enabling a calculation of the per-pixel dose in the presence of beam position jitter. By using {the Canny} edge detection algorithm~{\cite{Canny:1986}} we determined the borders of the area with unresponsive pixels and assigned a circle to it. The centre of this circle provides the pixel coordinates of the beam core for each image frame individually. {This procedure allows us to align the area with unresponsive pixels to the spatial beam distribution measured prior to the irradiation experiment and finally to reconstruct the number of photons delivered to individual ePix100a pixels with each pulse train as measured by the XGM. The per-pixel photon counts reconstructed for every ePix100a image frame and integrated over the duration of the irradiation results in the top left plot of figure~\ref{fig:dosedepth}.}

Finally, we used the GEANT4~\cite{pia:2003} based Monte Carlo simulation tool MULASSIS~\cite{Lei:2002a}, originally developed for dose and particle fluence analysis of shielding materials, to estimate the absorbed dose in different depths of the ePix100a sensor, taking {photoelectric} absorption in the different sensor materials into account. These are specifically the aluminum entrance window, the silicon bulk and the $\text{SiO}_2$ layers in the pixel structure as outlined in figure~\ref{fig:SensorCrosssection}. {The output of the simulation is normalized to the dose deposited per primary $\text{photon}$ and unit area $\text{cm}^2$, such that multiplying it with the measured per-pixel integrated photon number results in the total per-pixel dose.} The resulting spatial distribution of the dose deposited in the sensor at the depth of the $\text{SiO}_2$ structures and integrated over the time of all irradiation cycles is shown in the top {right} image of figure~\ref{fig:dosedepth}. The horizontal asymmetry of the distribution has its origin in the horizontal motion of the beam spot during the course of the irradiation. The vertical cuts through the sensor shown in figure~\ref{fig:dosedepth} illustrate the depth profile of the dose along the central column (bottom left image of figure~\ref{fig:dosedepth}) and row (bottom left image of figure~\ref{fig:dosedepth}).

While the region closest to the surface at the entrance window has received a maximum dose of $(1.3{\pm 0.2})\,\text{MGy}$ at column number $10$ and row number $9$, the dose deposited at the depth of the $\text{SiO}_2$ structures is reduced by a factor of $240$ to  $(5.4{\pm 0.7)}\,\text{kGy}$ due to absorption by the sensor material above. This corresponds to a total dose of $(180{\pm 13)}\,\text{MGy}$ delivered to the surface of the sensor when neglecting absorption in the sensor material and integrating the dose over the spot profile. The dose received by the ASIC is further reduced  by the shielding effect caused by the bump bonds. It can be estimated to $< 24\,\text{kGy}$ {for the ASIC areas located below the bump bonds and to $(670 \pm 140)\,\text{kGy}$ for the not shielded areas between the bump bonds. These numbers are translated to the surface of the ASIC, neglecting further structural details of the ASIC chip and absorption therein.}

\section{Experiment Results}
\label{sec:Experiment Results}
We characterized the pre- and post-irradiation performance of the ePix100a following the methodology as described in section~\ref{sec:Experiment Setup and Methodology}. The observed offset, noise and gain changes can be categorized into immediate and post-irradiation effects. We observed immediate effects during irradiation on time scales shorter than seconds or minutes. With post-irradiation effects we refer to changes of the detector performance on timescales of hours and days after the last irradiation cycle was completed. In the following sections we describe both effect categories in detail.

\subsection{Immediate Effects}
\label{sec:ImmediateEffects}
During irradiation with the high intensity FEL beam at an energy of $E_\text{Beam} = 1\,\mu\text{J}$, the pixels in the $20\,\text{pixels}\times20\,\text{pixels}$ large ROI {were} unresponsive. {We assume that this effect originates in the ASIC~\cite{Markovic:2014a, Carini:2012}. There is no obvious mechanism at the silicon sensor that we are aware of, which could cause the signal of individual pixels to drop to such low values as $130\,\text{ADU}$ on average. The mechanism by which this occurs has to be further investigated.}

{We would like to} emphasise that the observed effect {is of short duration} and only present during  irradiation. It does not lead to permanent damage: after irradiation, these pixels exhibit a dark signal saturating the detector’s dynamic range, but {their signal} recovers to an offset level comparable to the surrounding pixels within the following $48\,\text{hours}$. Furthermore, these pixels are fully functional during dark signal measurements, as will be shown in the following.

\begin{figure}[b]
\centering
\includegraphics[width=0.49\textwidth]{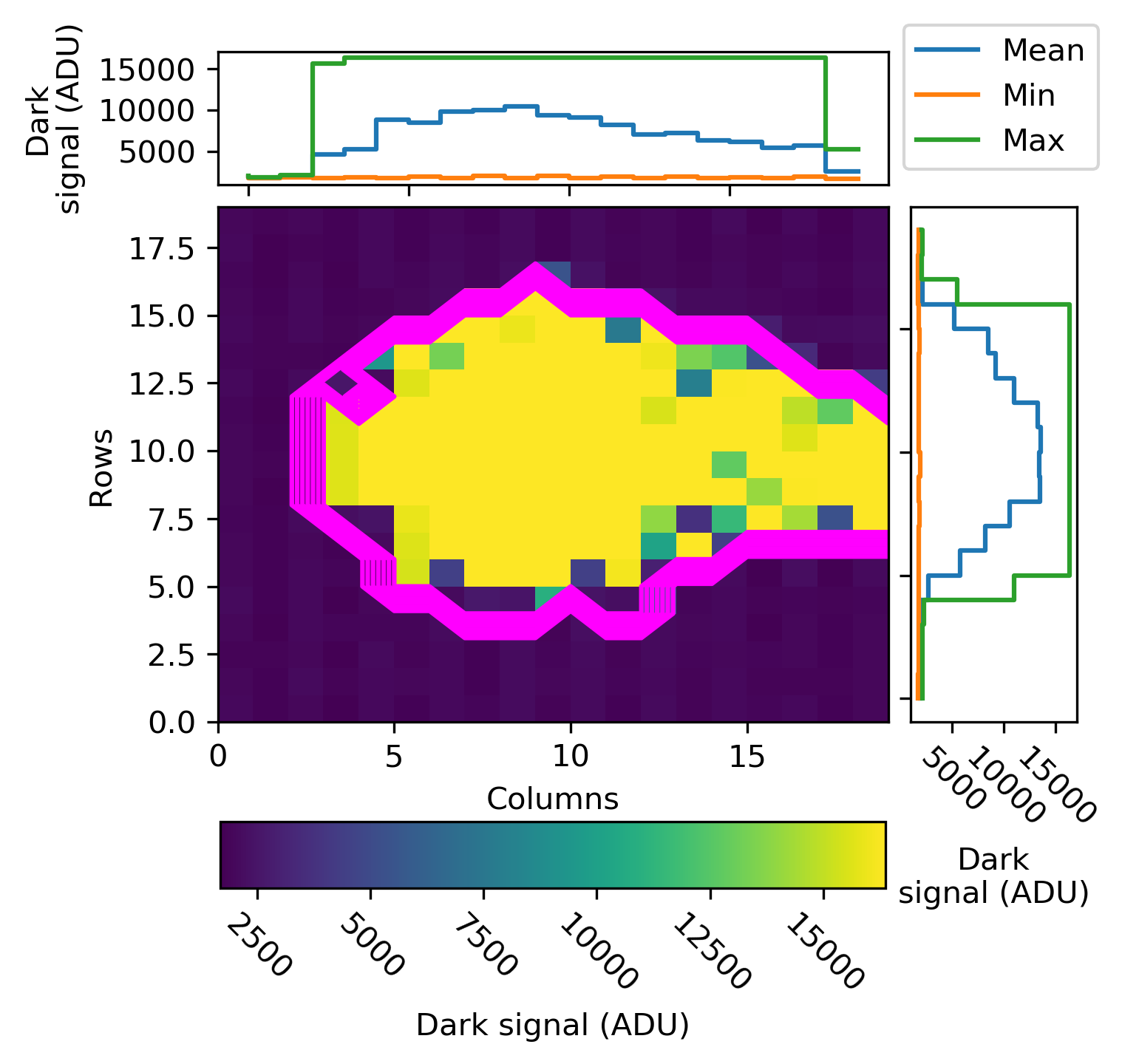}
\hfill
\includegraphics[width=0.49\textwidth]{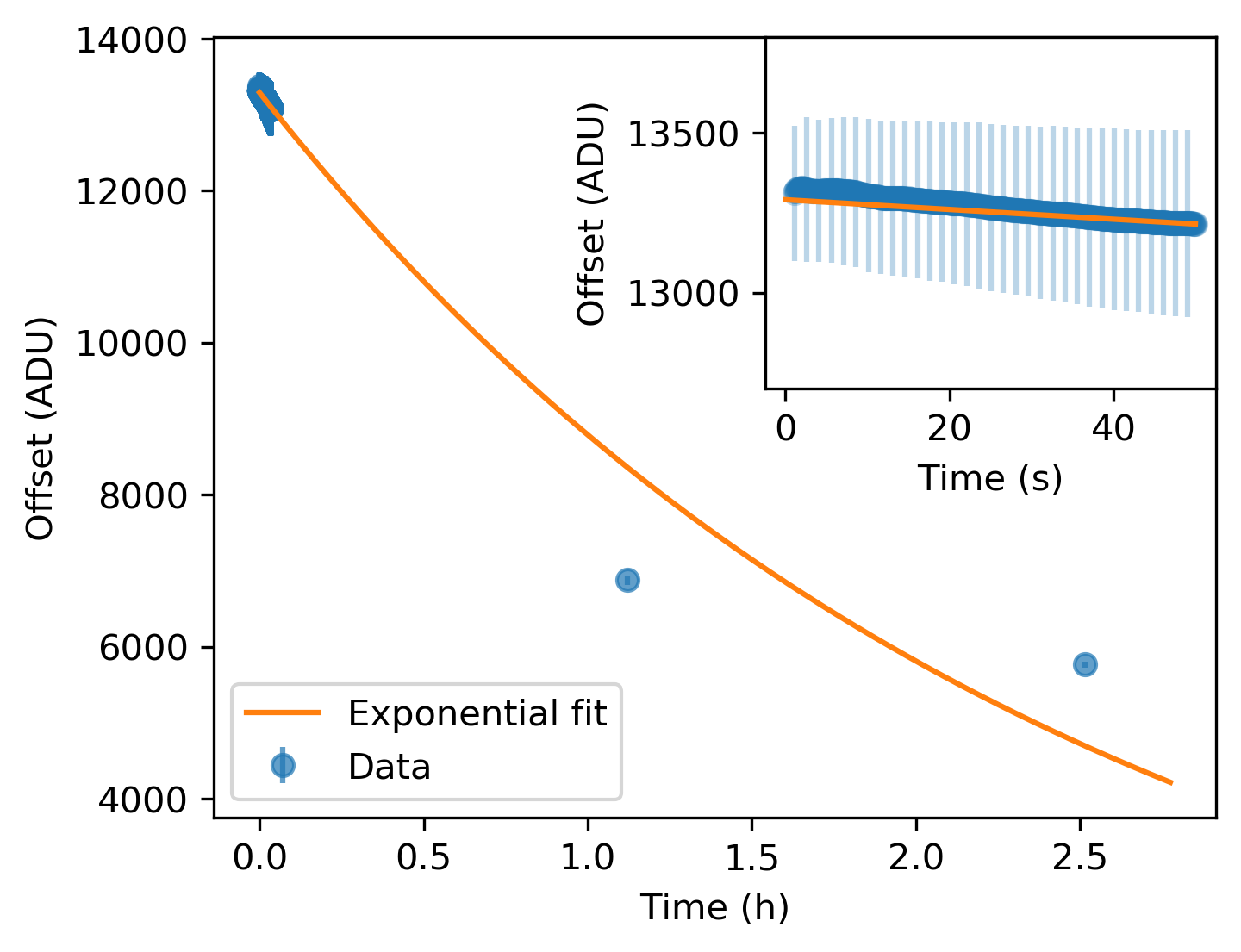}
\caption{\label{fig:SaturatedPx} Left: Image of the per pixel offset as observed in the ROI right after completion of the last irradiation cycle and after recovery from saturation. Pixels delivering an offset signal close or beyond the saturation threshold of the ADC have signal values $> 15500\,\text{ADU}$ (yellow color). Right: The evolution of the offset of those pixels located inside the magenta area {in the figure on the left is shown.} The evolution of the offset was monitored during two and a half hours after completion of the last irradiation cycle.}
\end{figure}

With an increasing number of irradiation cycles the number of pixels with an offset surpassing the upper end of the dynamic range of the ADC increased. The left part of figure~\ref{fig:SaturatedPx} shows the situation shortly after the last irradiation cycle was completed. The pixels shown in yellow were completely saturated. 

\subsection{Post-Irradiation and Long Term Effects}
\label{sec:Post-IrraditionEffects}
\subsubsection{Offset and Noise}
\label{sec:OffsetandNoise}
\begin{figure}[htbp]
\centering
\includegraphics[width=0.49\textwidth]{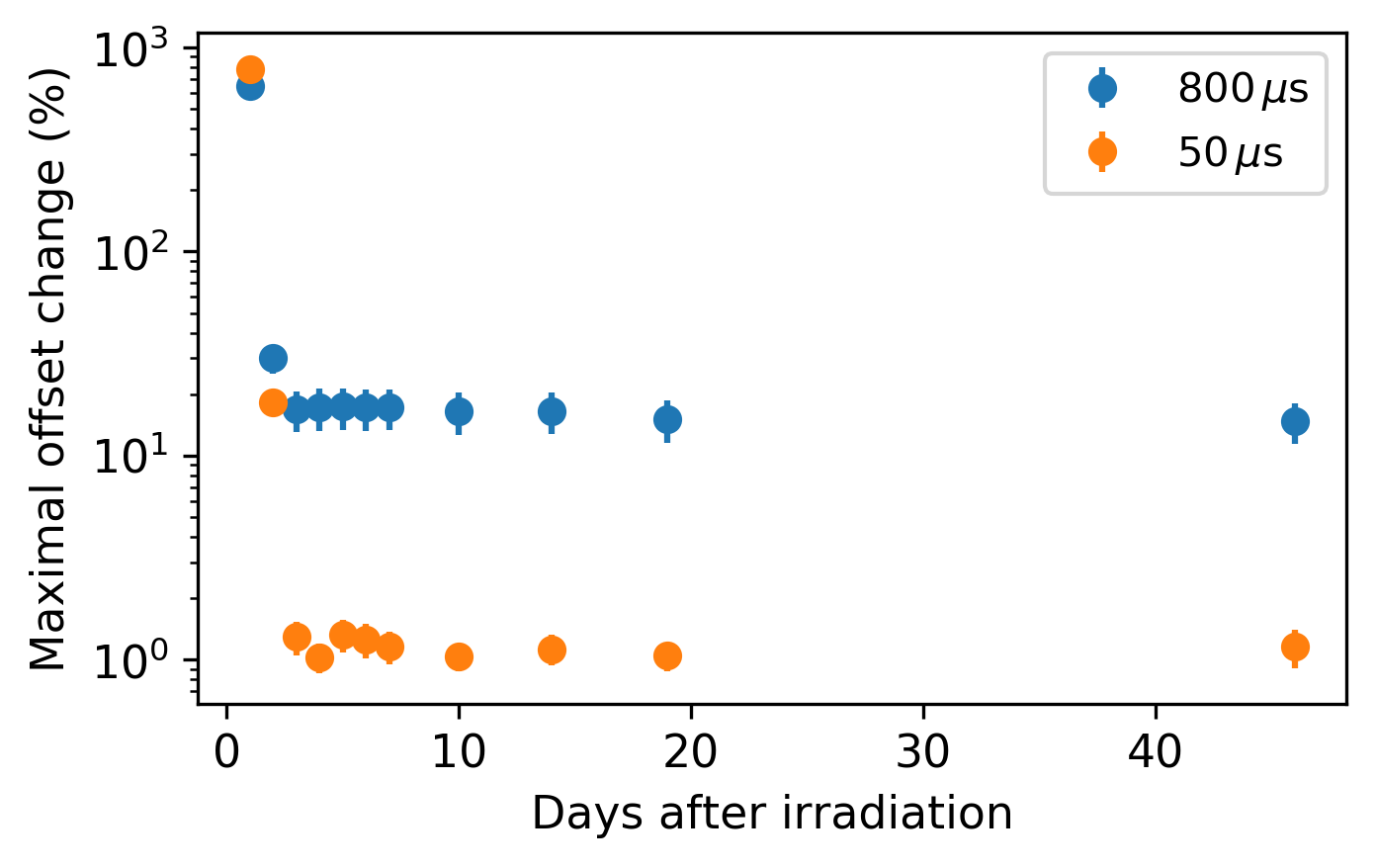}
\hfill
\includegraphics[width=0.49\textwidth]{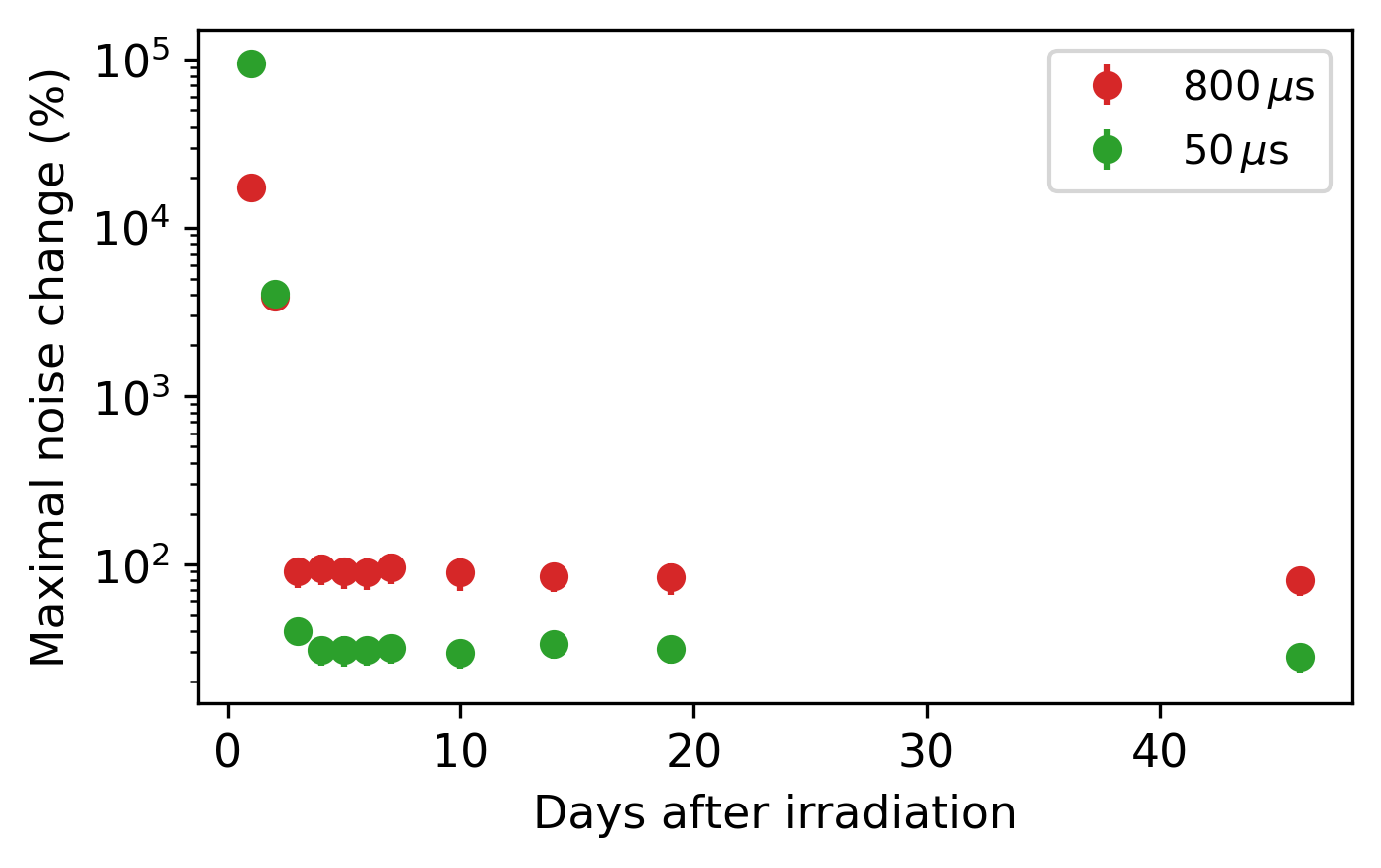}
\caption{\label{fig:maxChanges} Left: Evolution of the offset as observed in the pixel showing the highest relative offset {increase} for the two integration time settings, i.e. $t_\text{Int}=50\,\mu\text{s}$ and $t_\text{Int}=800\,\mu\text{s}$. The exponential decrease of the offset observed during the first day is followed by a stabilized state at higher offset values in comparison to the pre-irradiation level. Right: Relative change of the noise as observed during the days following the last irradiation cycle. The temporal behaviour of the noise mirrors the temporal evolution of the offset. {The average dose observed in the pixels showing the highest increase was $(4319 \pm 772)\,\text{Gy}$. The detector was kept at a temperature of $20\,^{\circ}\text{C}$ between the individual measurements.}}
\end{figure}

\begin{figure}
\centering
\includegraphics[width=0.5\textwidth]{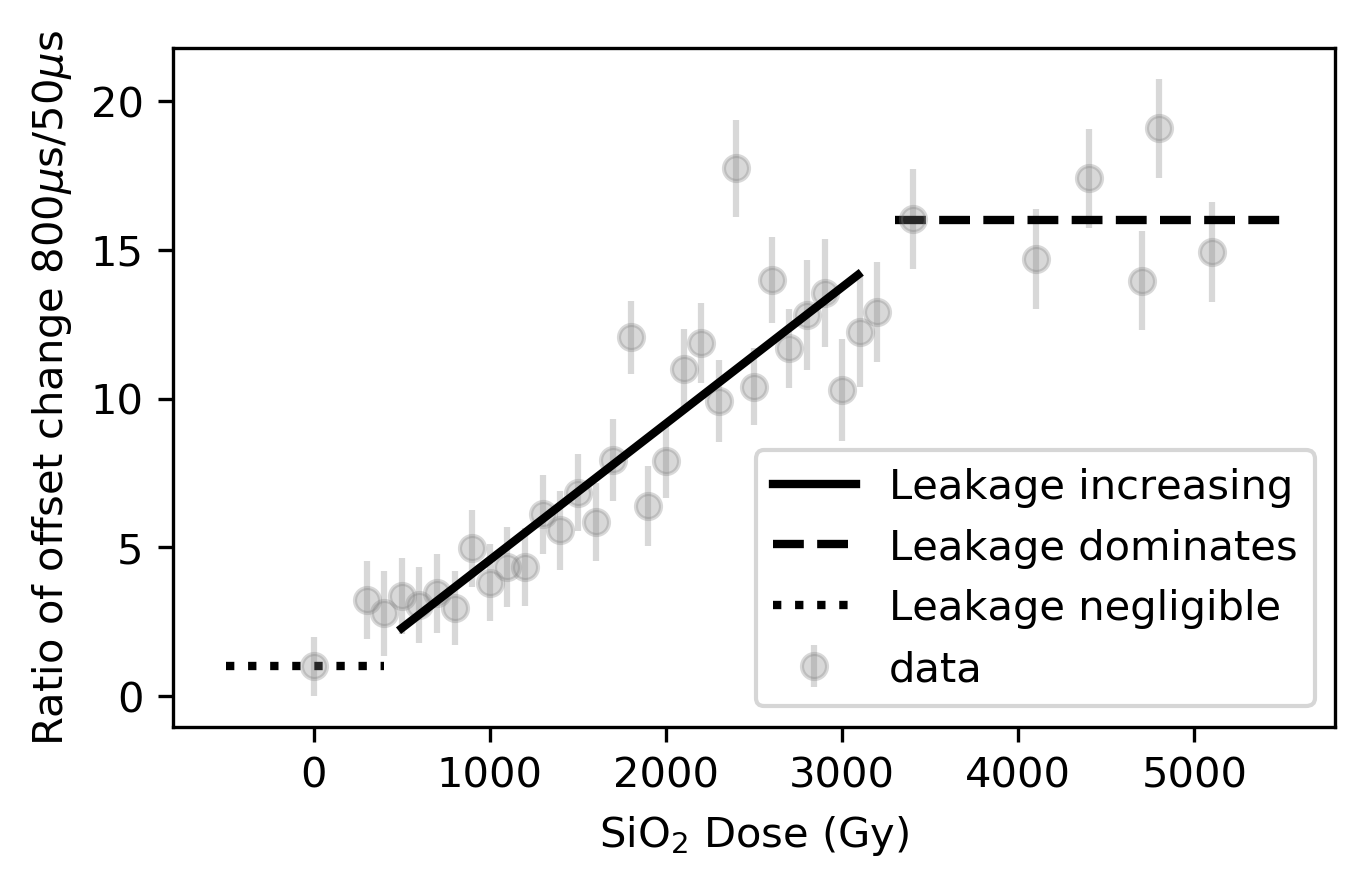}
\caption{Ratio of the offset change measured at $800\,\mu\text{s}$ and $50\,\mu\text{s}$ ({grey} dots). The black dashed line shows the scaling factor of $16$ expected from the ratio of the integration times when the leakage current dominates. If the leakage current is negligible the ratio equals $1$ (dotted line.) {The solid black line visualizes the region where the leakage current increases with the increasing dose.}}
\label{fig:IntegrationTimeRatio}
\end{figure}

During the first three hours following the last irradiation cycle the offset of individual pixels decreased exponentially with time with a decay coefficient of $-0.413\,\text{h}^{-1}$ as shown in the right part of figure~\ref{fig:SaturatedPx} for $t_\text{Int} = 800\,\mu\text{s}$. The offset stabilized three days after irradiation at a higher level of $1832\,\text{ADU}$ in comparison to the pre-irradiation level of $1762\,\text{ADU}$ (see figure~\ref{fig:maxChanges}). As is evident from figure~\ref{fig:maxChanges}, the scale of this effect measured $3$ days after irradiation remains the same also for the following measurements. In general we observe a larger offset and corresponding ENC for $t_\text{Int}=800\,\mu\text{s}$, when comparing pre- and post-irradiation conditions. Evaluating the offset change $46$ days after irradiation, yielded an offset increase by approximately $15\%$ for $t_\text{Int}=800\,\mu\text{s}$ and by $1\%$ for $t_\text{Int}=50\,\mu\text{s}$. The measured increase in offset scales linearly with the integration time, i.e. with a factor of $800\,\mu\text{s}/50\,\mu\text{s} = 16$, which is expected if the effect is predominantly due to an increase in dark current. As shown in figure~\ref{fig:IntegrationTimeRatio} the ratio of offset change for two integration times approaches the expected factor of $16$ when the leakage current increases, which happens for doses above $\approx 4000\,\text{Gy}$. {The figure can be divided into three regions, as illustrated by the dotted, solid and dashed lines. In first region (dashed line) the contribution of the leakage current is negligible at $0\,\text{Gy}$, hence no difference in the offset between shorter and longer integration time exists and the observed ratio is $1$. As the dose increases, the leakage current and in turn the ratio increases (second region, solid line). Finally, in the third region, the ratio approaches the ratio between the two integration times of $16$ (dashed line), which is expected when the leakage current dominates.}

\begin{figure}[htbp]
\centering
\includegraphics[width=0.47\textwidth]{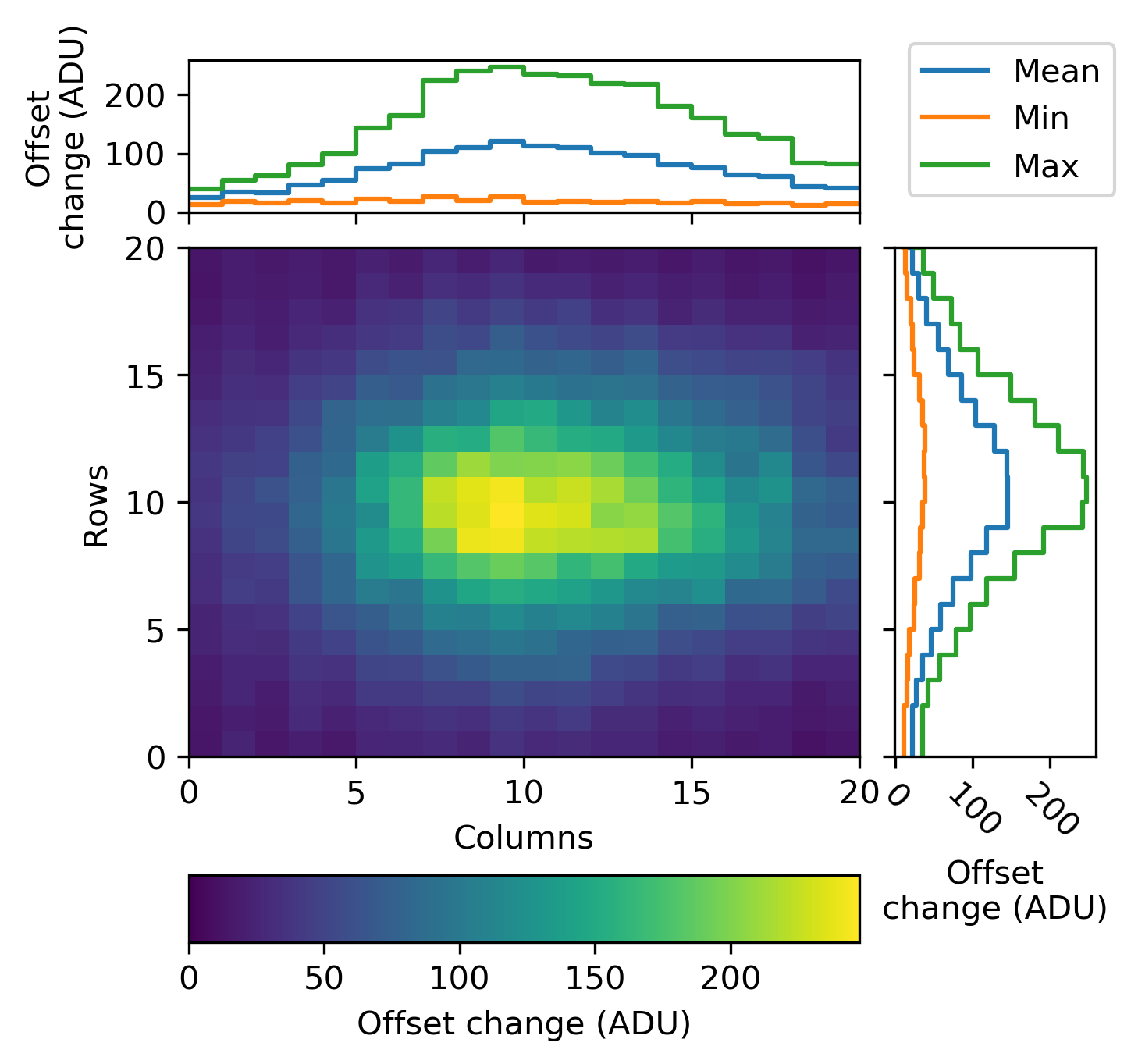}
\qquad
\includegraphics[width=0.47\textwidth]{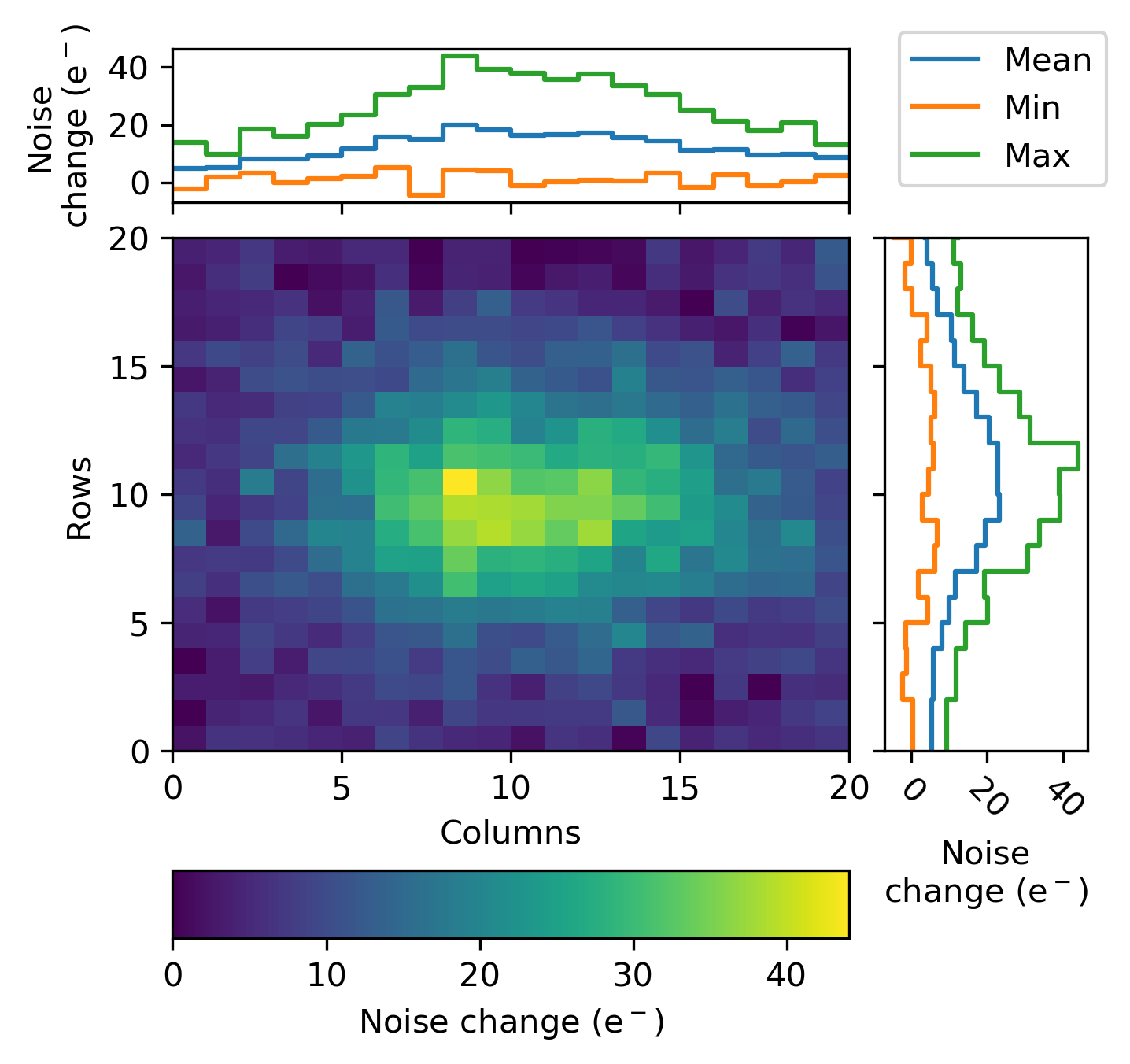}
\caption{Radiation-induced changes of the offset (left) and noise (right) for $t_\text{Int}=800\,\mu\text{s}$ evaluated 46 days after irradiation.}
\label{fig:DarkChangesABS}
\end{figure}

As shown in the right part of figure~\ref{fig:maxChanges}, the RMS noise observed in these pixels follows the same behaviour. While the noise at $t_\text{Int}=800\,\mu\text{s}$ has increased by 85\%, for $t_\text{Int} = 50\,\mu\text{s}$ the increase is at the level of 30\%. {The maximal increase of the offset and noise 46 days after irradiation is measured for the average dose of $(4683 \pm 659)\,\text{Gy}$.} The spatial distribution of the induced offset (left) and noise (right) changes is shown in figure~\ref{fig:DarkChangesABS} for $t_\text{Int}=800\,\mu\text{s}$.

Since different pixels within the ROI have received a different dose, we can evaluate the change of the offset and RMS noise between pre- and post-irradiation conditions depending on the dose, when assuming that the pixels inside the ROI react similarly to radiation induced damage. Since the design of the pixels is the same {and the observed pixel-to-pixel variations of the dark current, noise and amplification are $\approx 3\,\%$, we consider this assumption to be justified.}

\begin{figure}[t]
\centering
\includegraphics[width=0.49\textwidth]{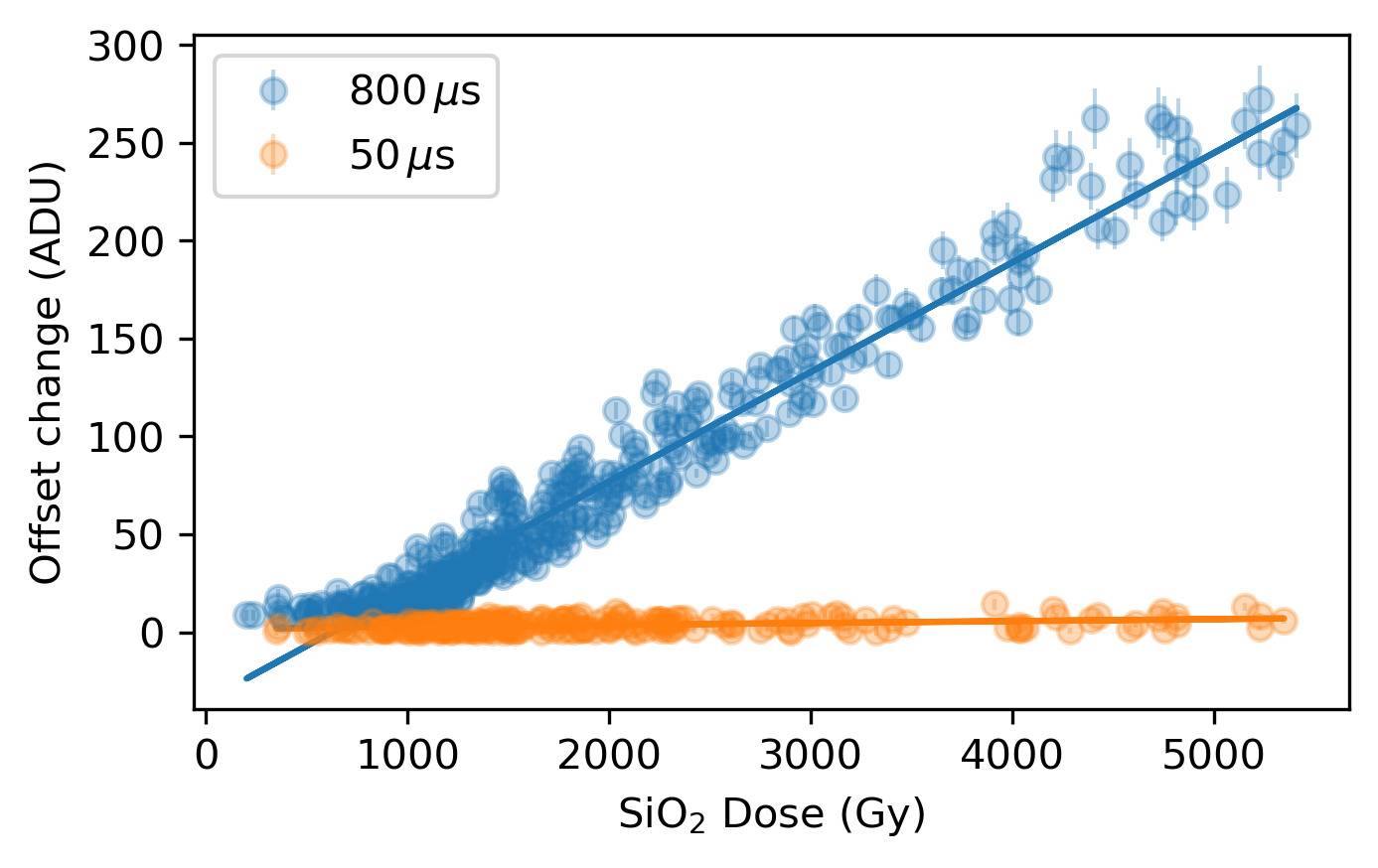}
\hfill
\includegraphics[width=0.49\textwidth]{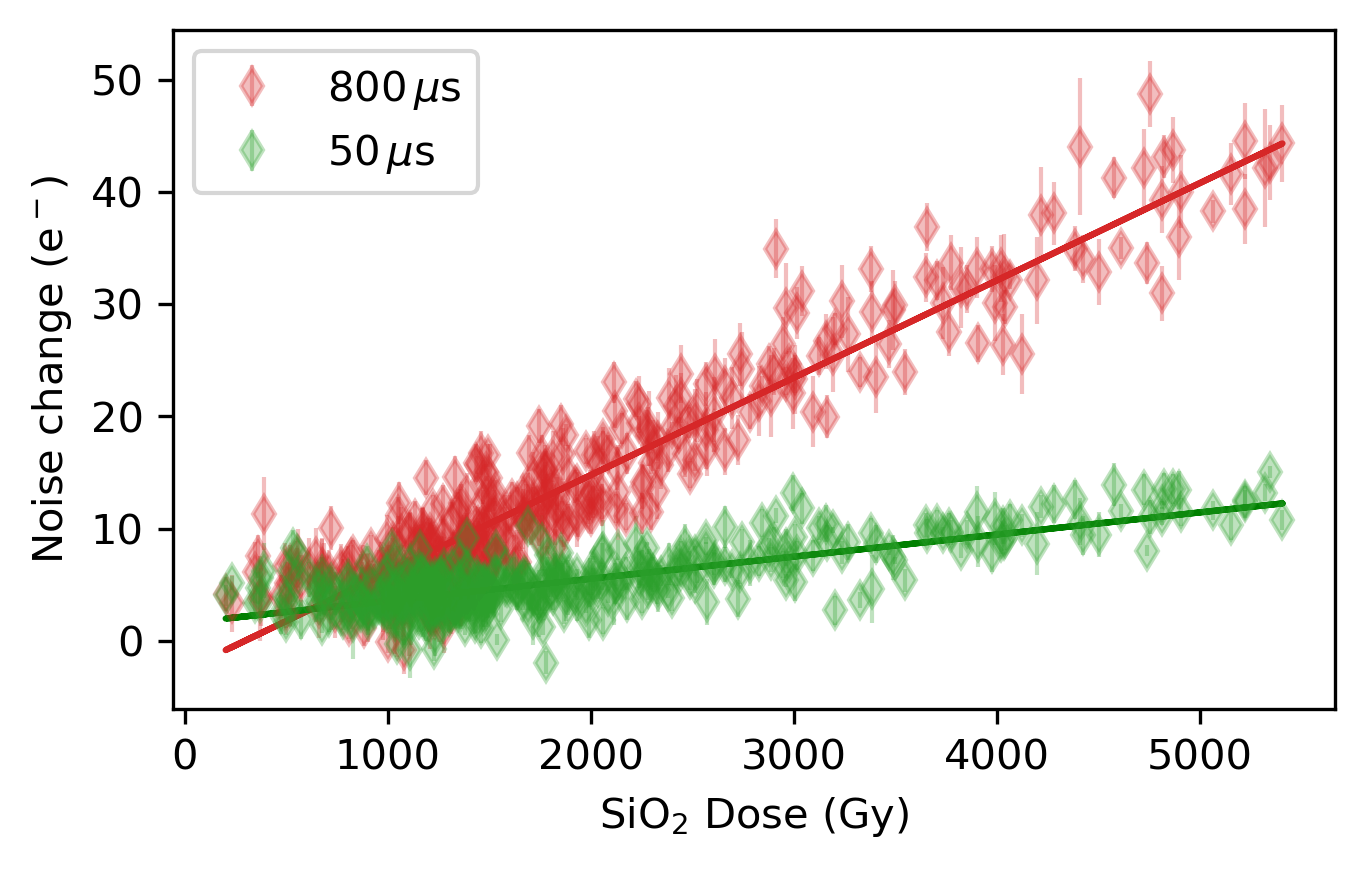}
\caption{\label{fig:Dose_vsChanges} Left: The change of the offset observed in different pixels of the ROI {46 days after irradiation} when comparing pre- and post-irradiation conditions depending on the accumulated dose at the depth of $\text{SiO}_2$ layers is shown for $t_\text{Int}=50\,\mu\text{s}$ and $t_\text{Int}=800\,\mu\text{s}$. Right: Dependency of the ENC observed in different pixels of the ROI on the accumulated dose in units of electrons for the same integration time setting.}
\end{figure}

Figure~\ref{fig:Dose_vsChanges} shows the offset (left) and noise (right) changes depending on the absorbed dose measured at the depth of the $\text{SiO}_{2}$ interface 46 days post-irradiation. Here again the influence of the longer integration time is clearly visible. The slope derived from fitting a linear function to the data yields an offset and ENC change rate of $(56.0\pm0.6)\,\text{ADU}/\text{kGy}$ and $(8.7\pm0.1)\,\text{e}^-/\text{kGy}$ for $t_\text{Int}=800\,\mu\text{s}$  and $(1.0\pm0.2)\,\text{ADU}/\text{kGy}$ and $(2.0\pm0.1)\,\text{e}^-/\text{kGy}$ for $t_\text{Int}=50\,\mu\text{s}$, respectively. 
If absorption of the radiation in silicon is neglected when calculating the dose, the offset and ENC change rate yields $(235.9\pm2.6)\,\text{ADU}/\text{MGy}$ and $(37.4\pm0.6)\,\text{e}^-/\text{MGy}$ for $t_\text{Int}=800\,\mu\text{s}$  and $(4.2\pm0.6)\,\text{ADU}/\text{MGy}$ and $(8.3\pm0.4)\,\text{e}^-/\text{MGy}$ for $t_\text{Int}=50\,\mu\text{s}$.
The maximum observed increase of the offset reduces the available dynamic range of the detector by approximately $2\%$ for $t_\text{Int}=800\,\mu\text{s}$ and $\approx 0.1\%$ for $t_\text{Int}=50\,\mu\text{s}$. 

\subsubsection{Gain and Energy Resolution}
\label{subsec:Gain&EnergyResolution}
We took post-irradiation {copper} fluorescence flat-field data approximately one and a half hours after completion of the last irradiation cycle. At that time most of the pixels in the central part of  the {region of interest} were still in saturation, thus not capable of detecting charge created by a photon interaction. Later, calibration measurements (with the detector in stabilized post-irradiation state) were not possible due to time-constrained access to the instrument. In order to characterise the gain and energy resolution, we therefore used pixels located in the periphery of the saturated area to compare the pre- and post-irradiation performance.

\begin{figure}[t]
\centering
\includegraphics[width=0.49\textwidth]{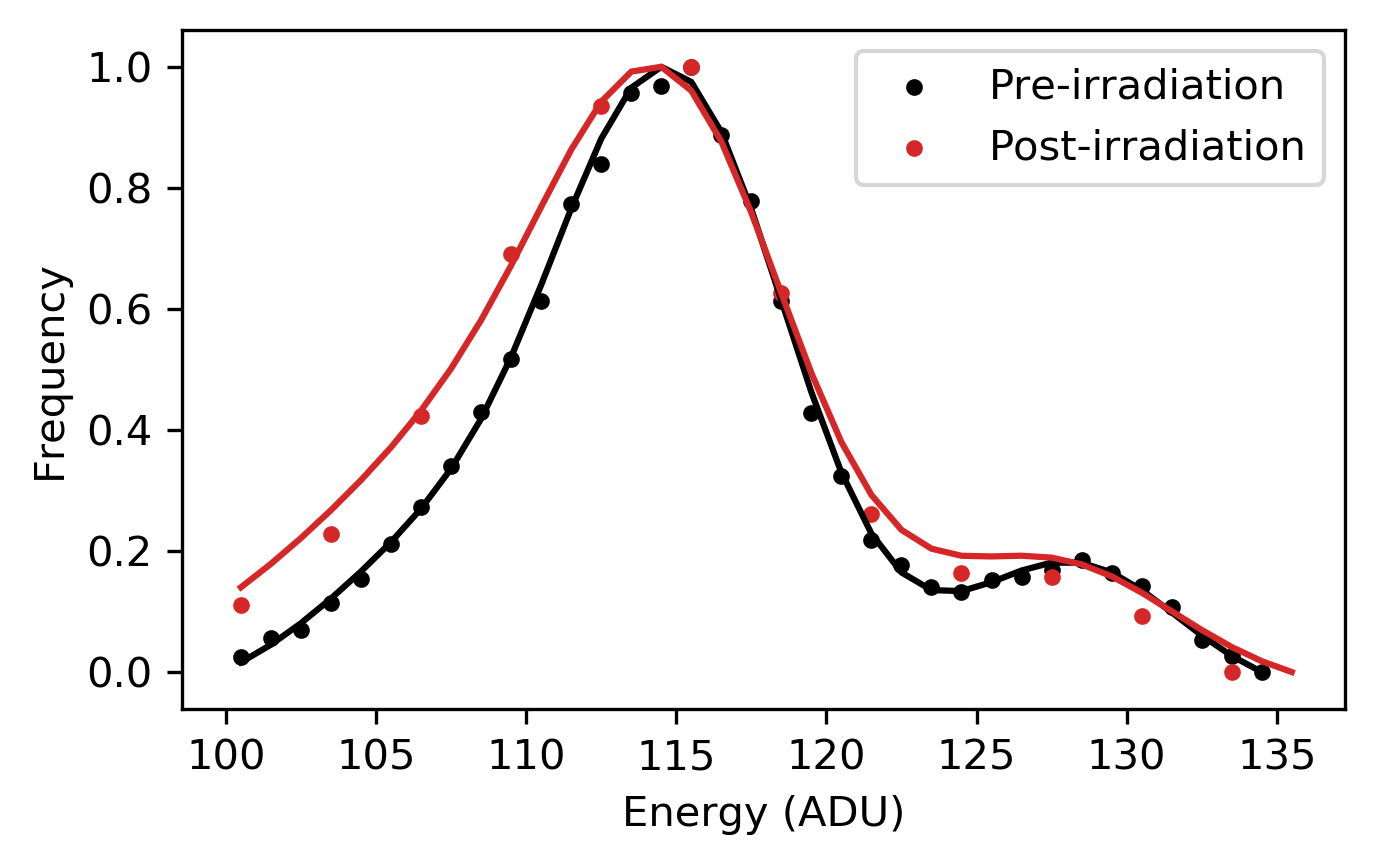}
\hfill
\includegraphics[width=0.49\textwidth]{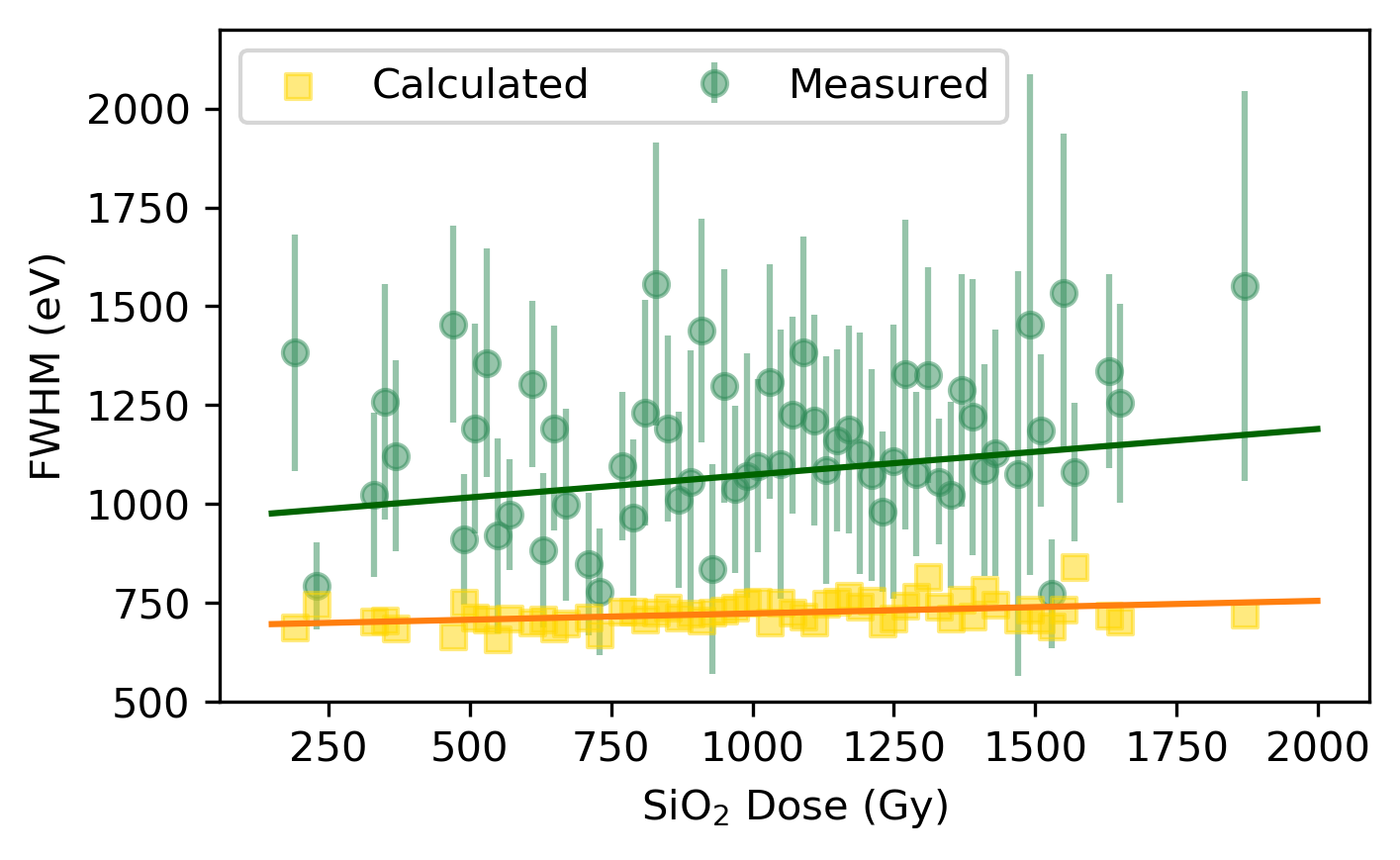}
\caption{\label{fig:Cu_spec} Left: The spectral distribution of the {$\text{Cu-K}_{\alpha}$} photons and $9\,\text{keV}$ photon peak as detected by pixels in the periphery of the {region of interest} before irradiation (black) and after irradiation (red). Right: Relation of the {$\text{Cu-K}_{\alpha}$ line} peak width (assessed as FWHM) to the dose absorbed in the $\text{SiO}_2$ layer. Values extracted from measurement are marked with green dots, while the yellow squares shows the expected FWHM calculated from the noise in pixels.}
\end{figure}

The left part of figure~\ref{fig:Cu_spec} shows a comparison of the measured spectrum of the { $\text{Cu-K}_{\alpha}$} lines and the $9\,\text{keV}$ line resulting from the photon beam before (black) and after irradiation (red) calculated from {the peripheral} pixels in the {region of interest}. The FWHM of the lines is larger after irradiation and the lines have a more pronounced low energy tail. 

Figure~\ref{fig:Cu_spec} on the right shows the dependency of the peaks' FWHM on the dose accumulated by each individual pixel. We find an increase of the FWHM of $100\,\text{eV}$ per $866\,\text{Gy}$. Following the relation of the intrinsic resolution limit of a semiconductor detector $\Delta E \propto \sqrt{FWE}$, with the Fano factor {$F = 0.120$, the photon energy $E$ and the energy required to create an electron hole pair {$W=3.66\,\text{eV}$} \cite{Lowe:2007a, Mazziotta:2008a} },  we would expect the behaviour indicated by yellow squares (labeled as "Calculated"). The increase of the FWHM values follows the same slope (within the estimated errors) as the calculated intrinsic resolution values. The intercept values of the linear models are separated by ${184}\,\text{eV}$ (best case scenario including $1\,\sigma$ uncertainties), which is approximately consistent with the increased mean noise value measured in the {peripheral pixels of the region of interest} after irradiation. These results suggest, that the observed broadening of the {$\text{Cu-K}_{\alpha}$ line} is driven by the radiation-induced noise increase.
It is to note that in order to convert ADU values to eV/$\text{e}^-$ units a pre-irradiation absolute gain value of $g=({70.2}\pm0.5)\,\text{eV}/\text{ADU}$ was used. 

\begin{figure}[ht]
\centering
\includegraphics[width=0.49\textwidth]{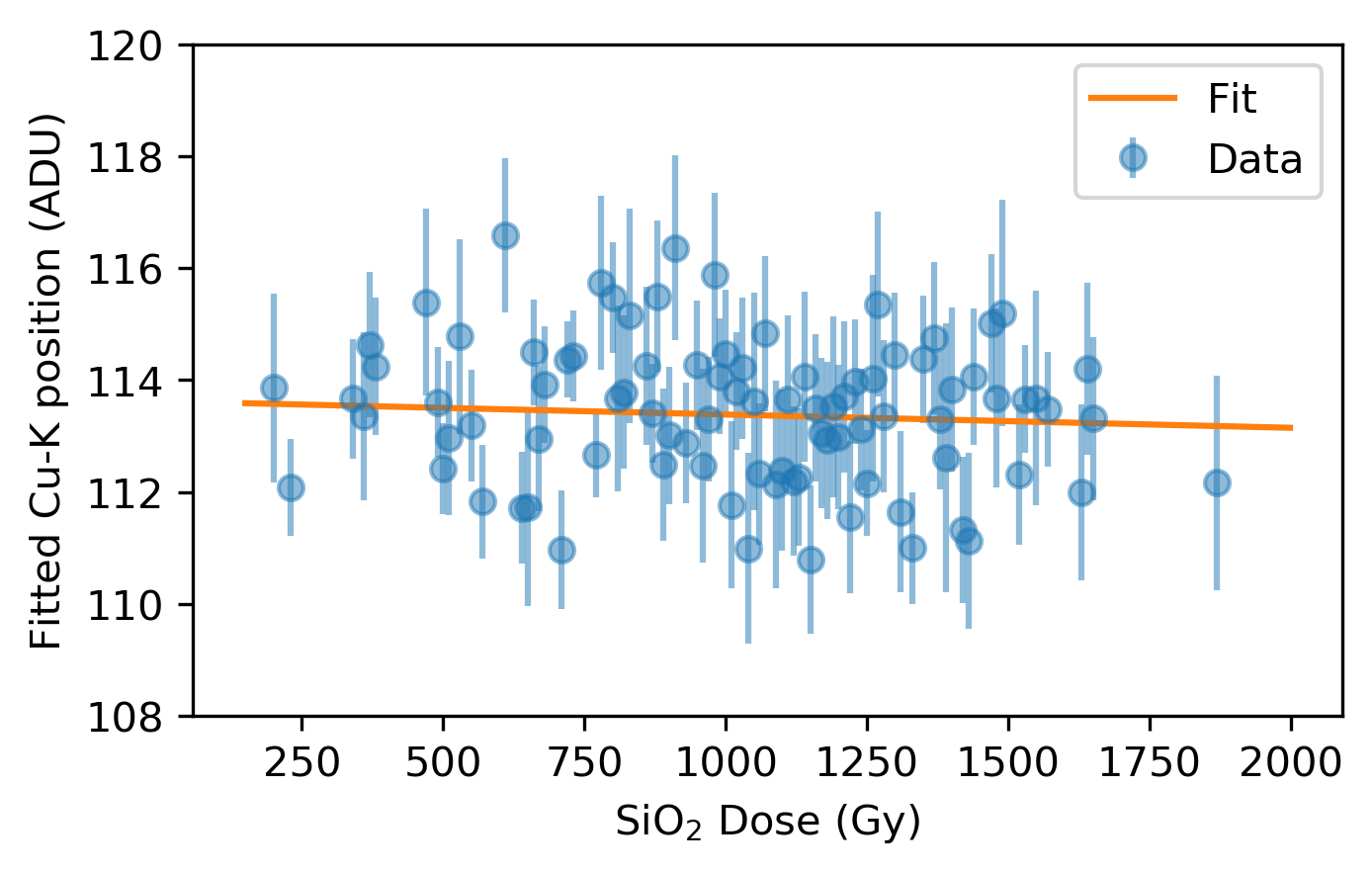}
\caption{\label{fig:PeakPos} Position of the {$\text{Cu-K}_{\alpha}$} line in dependency of an accumulated dose in different pixels. The line position was determined by fitting our Gaussian model to the spectra.}
\end{figure}

Before irradiation we find $(114.75\pm0.07)\,\text{ADU}$ for the position of the { $\text{Cu-K}_\alpha$} line blend. Figure~\ref{fig:PeakPos} shows the position of the { $\text{Cu-K}_{\alpha}$} line extracted from single pixel spectra of irradiated pixels in dependence of the absorbed dose. Applying a linear model to the data yields a slope which is consistent with $0$, indicating no change of the {$\text{Cu-K}_{\alpha}$ line} position with dose. The { $\text{Cu-K}_\alpha$ line} after irradiation are located at $(113.62\pm0.37)\,\text{ADU}$ as we determined from the linear model.

Since the flat-field calibration measurement contained only pixels with an absorbed dose up to a maximum of $2\,\text{kGy}$, the gain behaviour for the most irradiated pixels could not be studied. In order to study the gain behaviour of these pixels, we performed a charge injection scan with the current sources implemented in each pixel $240$ days after irradiation. During the linearity scan an increasing signal is injected into a preamplifier with an internal 10-bit pulser, thus simulating a charge created by photon interactions in the sensor material. For our measurement, 1024 steps of the pulser were used to scan the full dynamic range of the detector. 

\begin{figure}[htbp]
\centering
\includegraphics[width=0.46\textwidth]{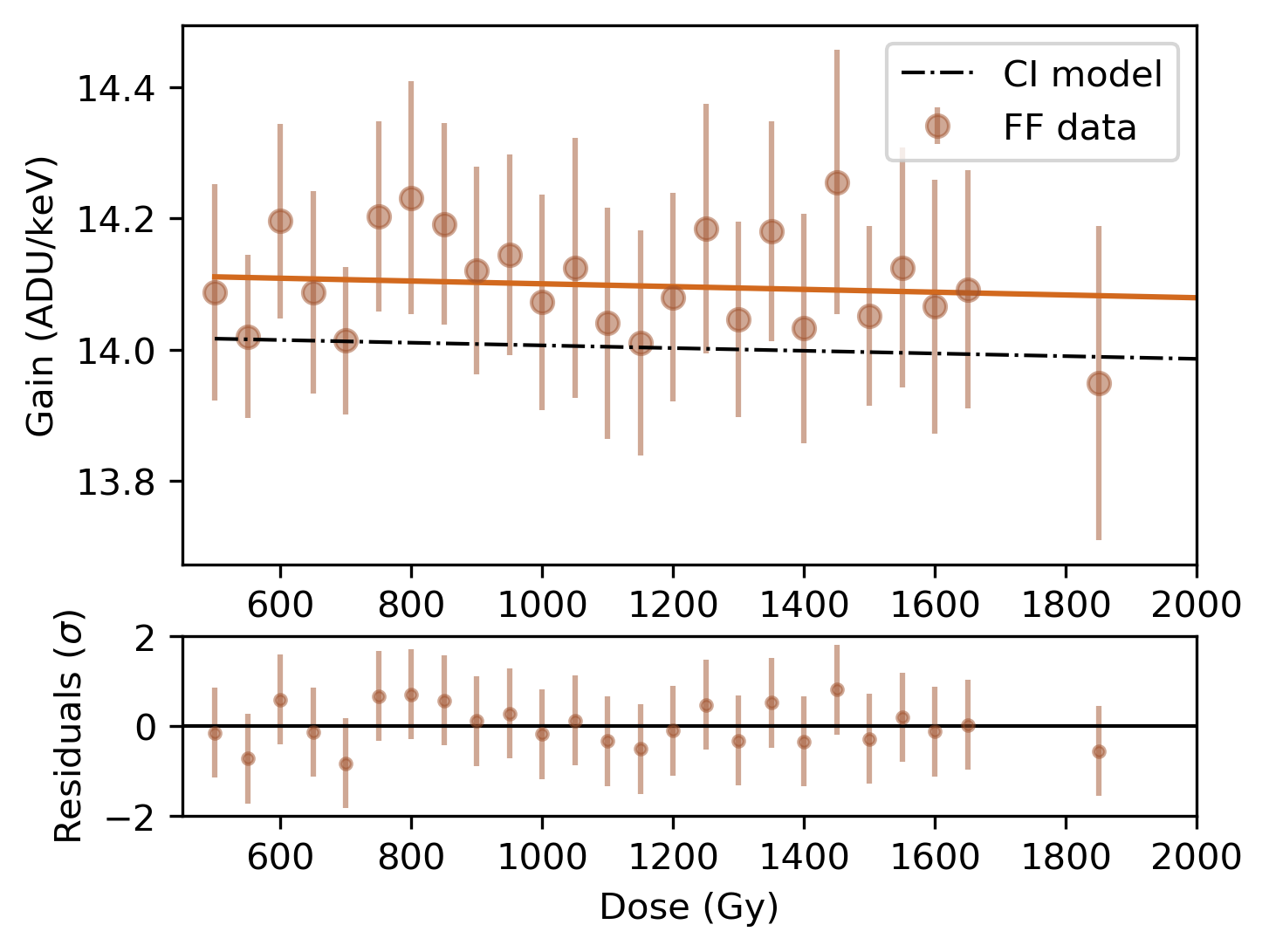}
\qquad
\includegraphics[width=0.46\textwidth]{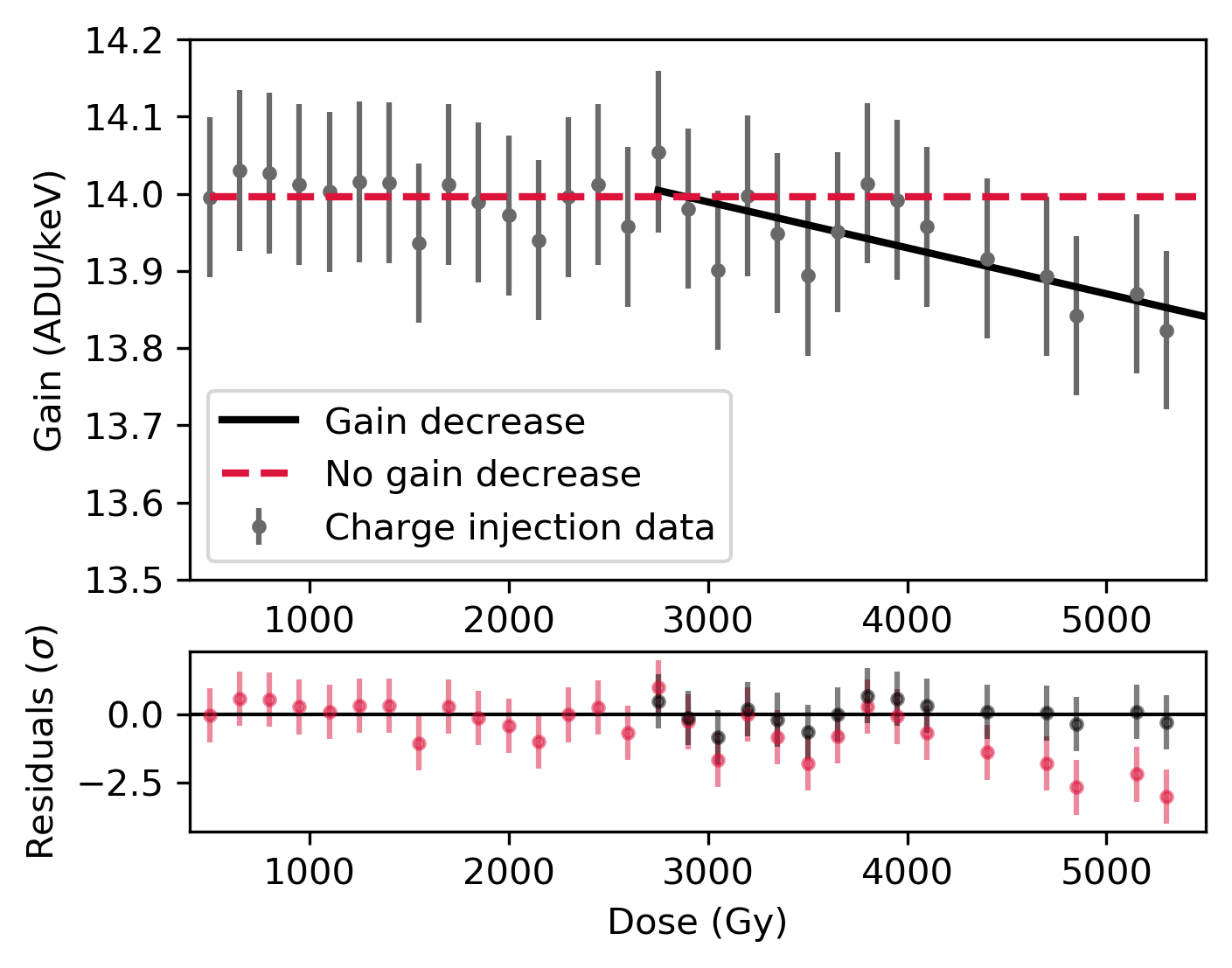}
\caption{\label{fig:Gains} Left: Gain estimated from the {$\text{Cu-K}_{\alpha}$} fluorescence data depending on dose. The black line shows a model derived from charge injection data (marked as "CI model") estimated for the absorbed dose in the range of $500\,\text{Gy}$ and $2600\,\text{Gy}$. Right: Gain calculated from the internal charge injection data in relation to the absorbed dose. { Two regions were fitted separately, lower dose region ($\leq 2600\,\text{Gy}$) where no gain decrease is observed and higher dose region ($> 2600\,\text{Gy}$) showing a weak gain decrease. Residuals (red dots) of the constant function extended to the whole deposited dose range show a systematic deviation from the charge injection data of up to $-3\sigma$ for gain values above $\approx 4\,\text{kGy}$. Hence, demonstrating a weak gain decrease.}}
\end{figure}

The observed gain values derived from {$\text{Cu-K}_{\alpha}$} fluorescence data are shown in dependency of the absorbed dose in the left part of figure~\ref{fig:Gains}. The x-axis range is limited to dose values below $2\,\text{kGy}$, as the pixels which have seen a higher dose were still saturated at the time the flat-field data was taken (see section \ref{sec:ImmediateEffects}). The slope $(-2\pm 9)\times 10^{-5}\,\text{ADU}\,\text{keV}^{-1}\text{Gy}^{-1}$ derived from a linear model fitted to the data is consistent with zero, indicating that the gain is not changing significantly up to a dose of $\approx 2\,\text{kGy}$. This behavior changes as soon as higher dose levels are reached. If we take the charge injection data covering the range between {\bf $2750\,\text{Gy}$} and $5500\,\text{Gy}$ into consideration (right part of figure~\ref{fig:Gains}), we find the gain decreasing with the rate of {$(-6\pm 3)\times 10^{-5}\,\text{ADU}\,\text{keV}^{-1}\text{Gy}^{-1}$} as indicated by the black {solid} line in the right panel of {f}igure~\ref{fig:Gains}. {Gain values for lower deposited dose, i.e. deposited dose $\leq 2600\,\text{ADU}$ do not show a decrease as shown by the dashed red line. 
The significance of the gain decrease with higher doses is visualised by residuals of a constant function fitted to charge injection data up to a dose of $2600\,\text{Gy}$ and extended to the full range of deposited doses. As shown by the red dots in residuals plot in figure~\ref{fig:Gains}, the gain values for doses above $4\,\text{kGy}$ deviate up to $-3\sigma$, thus indicating a weak gain decrease. On the other hand, the residuals visualized by black dots corresponding to the linear fit to gain values for doses above $2750\,\text{Gy}$ show a very good agreement between the measured values and the fitted function.} 


\section{Interpretation}
We attribute the measured increase of the leakage current and consequently higher offset after irradiation to radiation induced damage in the sensor by a mechanism discussed by Schwandt et al.~\cite{Schwandt:2012a}. The generation of $\text{e}^{-}\text{-- h}$ pairs close to the $\text{Si--SiO}_2$ interface and build up of positive charge lead to high electric fields near the interface causing changes to the depletion boundary.
Coulomb repulsion happening between positively doped boron implantation and positive charges accumulating near the oxide layer might result in a shrinking of the boron implanted region on its edges, thus exposing the metal contact and allowing the depleted area to extend to the region close to the metal contact. Potentially, the bending of the depletion boundary can reach the edges of the metal contact and thus increase the electron leakage current. 
When the generation of new charge carriers is interrupted, the recombination process dominates and the leakage current will decrease exponentially as observed and shown in the right part of figure~\ref{fig:SaturatedPx}. 

The effects on the gain were found to be pronounced at higher doses and are visible in the charge injection scan data. This indicates a radiation effect on the readout electronics in the ASIC, as the sensor-induced signal does not significantly contribute to those measurements. We see different mechanisms which could lead to such an observed gain decrease. However, presently, it is not fully understood which of these mechanisms contribute to the gain change observed in the ASIC.

To explain the gain changes and pixel saturation, both of which occur on an ASIC-level in more detail, a device simulation based on the specific design of the ePix100a sensor would be required, which is beyond the scope of this work.

\section{Detector Lifetime Estimate}
The radiation-induced damage presented in the previous section can be used to estimate the lifetime of the detector depending on the beam energy used during experiments and limits for the measurement time beyond which the performance of the detector will significantly degrade. The estimates presented here are based on the extrapolation of the measured relationship of the induced damage and dose absorbed in the $\text{SiO}_2$ layer. 

\begin{figure}
\centering
\includegraphics[width=0.49\textwidth]{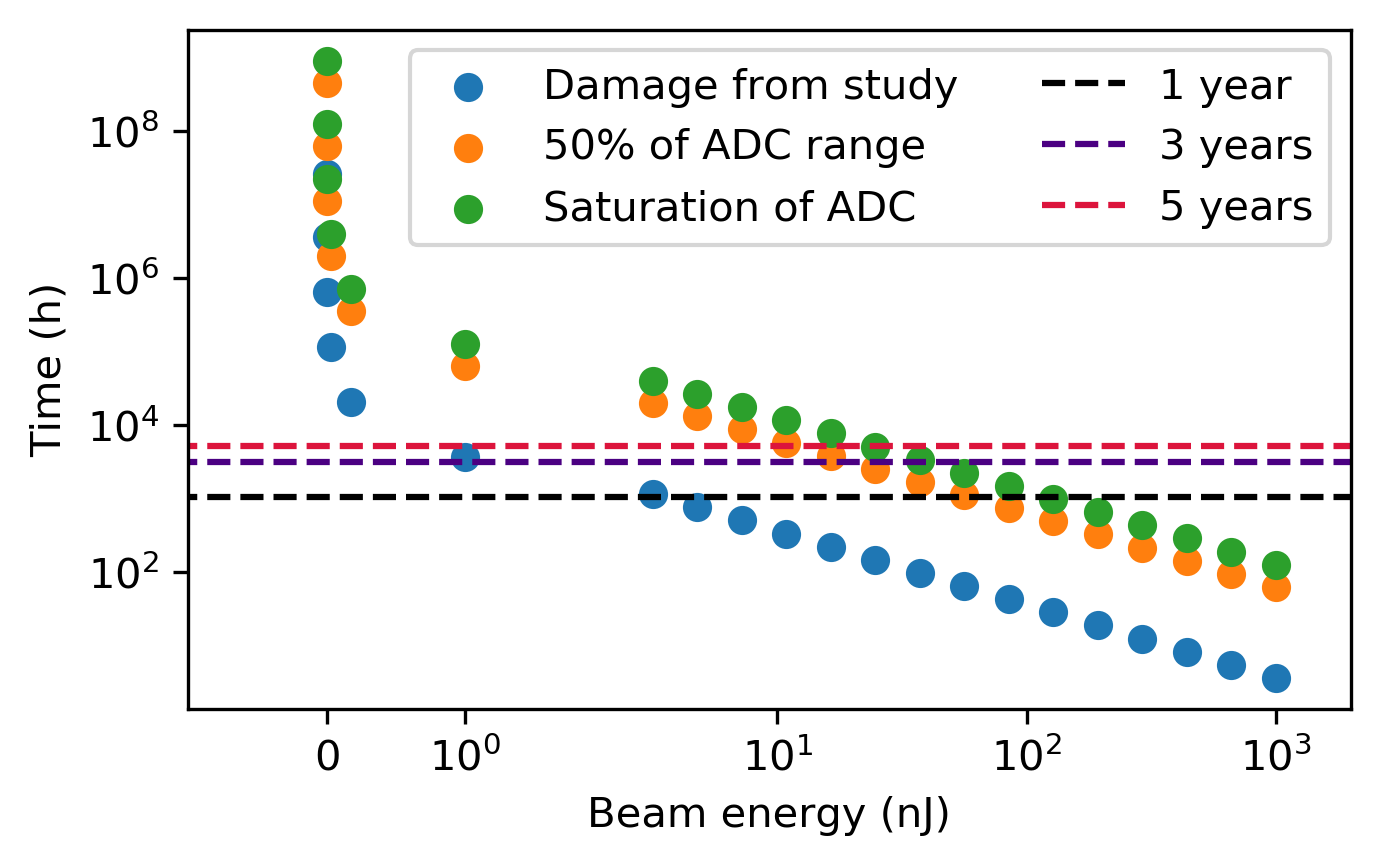}
\hfill
\includegraphics[width=0.49\textwidth]{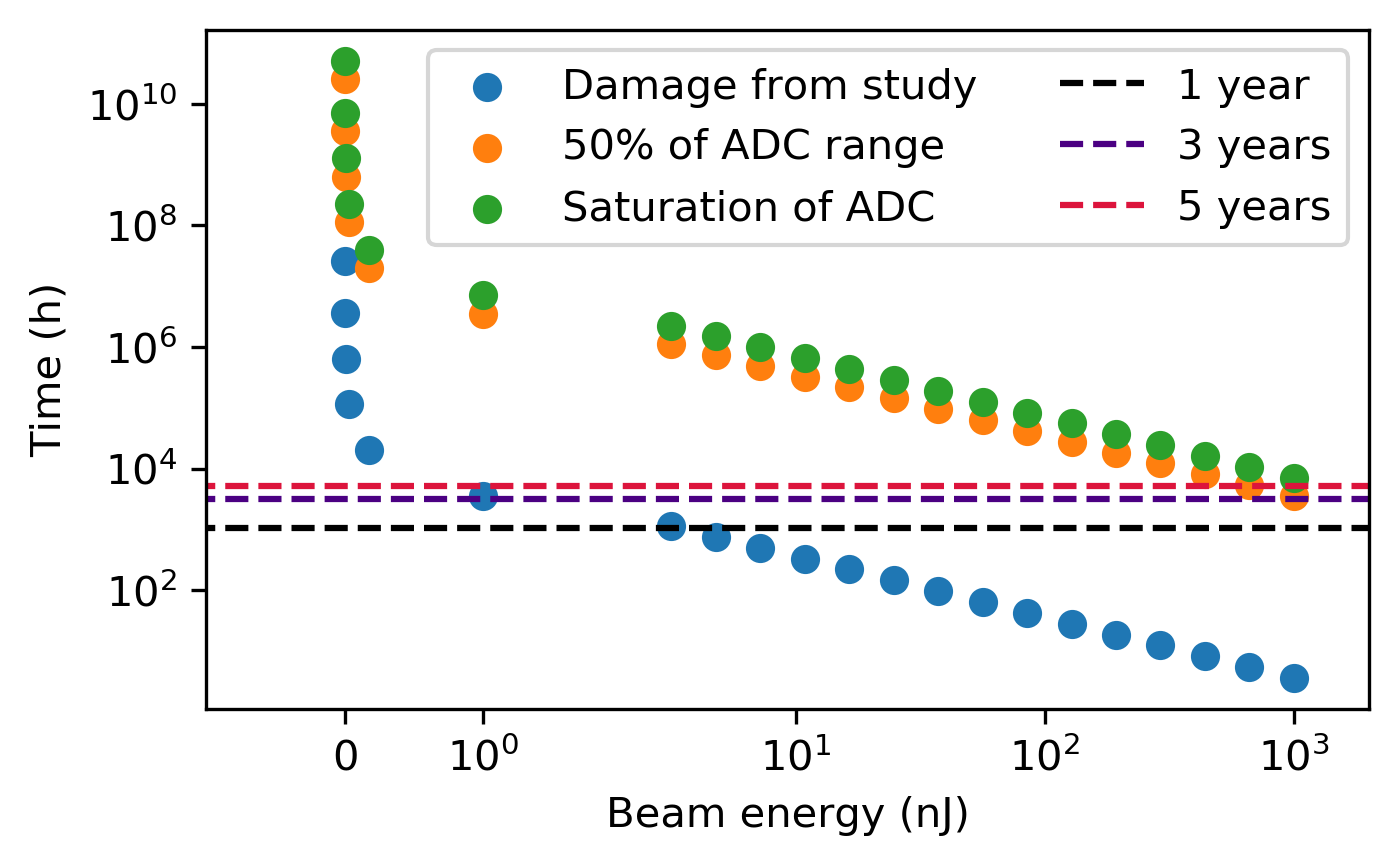}
\caption{Estimate of time needed to reach specific level of the dynamic range reduction at a certain beam energy. Three scenarios are shown; the reduction level observed during this radiation hardness study (blue dots), reduction to $50\,\%$ of the initial dynamic range (orange dots) and complete saturation of the ADC by the leakage current (green dots). The left plot shows the estimate for $t_\text{Int} = 800\,\mu\text{s}$ and the right plot is for $t_\text{Int} = 50\,\mu\text{s}$.} 
\label{fig:LifetimeRange}
\end{figure}

Figure~\ref{fig:LifetimeRange} illustrates the time needed for a specific dynamic range reduction depending on the beam energy. A reduction of the dynamic range by $50\%$ can be expected at a dose of ca. $(131{\pm 18)}\,\text{kGy}$ for $t_\text{Int}=800\,\mu\text{s}$ and at ca. $(7.4{\pm 1.0)}\,\text{MGy}$ for $t_\text{Int}=50\,\mu\text{s}$ absorbed in a maximally irradiated pixel{. Assuming a beam spatial distribution similar to the one used during this radiation damage experiment, i.e. the most irradiated pixel receiving $1\,\%$ of the total beam energy, the dose per $20\,\text{pixels}\times20\,\text{pixels}$ area} amounts to ca. $(13{\pm 1)}\,\text{MGy}$ for $t_\text{Int}=800\,\mu\text{s}$ and $(740{\pm 64)}\,\text{MGy}$ for $t_\text{Int}=50\,\mu\text{s}$. Saturation of the ADC dynamic range will occur at $(262{\pm 36)}\,\text{kGy}$ ($t_\text{Int}=800\,\mu\text{s}$), respectively at $(14.8{\pm 2.0)}\,\text{MGy}$ ($t_\text{Int}=50\,\mu\text{s}$), i.e. at $(26{\pm 2)}\,\text{MGy}$ ($t_\text{Int}=800\,\mu\text{s}$) and at $(1.48{\pm 0.13)}\,\text{GGy}$ ($t_\text{Int}=50\,\mu\text{s}$) of the total absorbed dose in ROI. 
The left panel of figure \ref{fig:LifetimeRange} shows three exemplary cases for the expected dynamic range behaviour; dynamic range reduction as observed in this radiation hardness study (blue dots), extrapolated loss of $50\%$ of the ADC range (orange dots) and saturation of the ADC dynamic range due to the leakage current (green dots) for $800\,\mu\text{s}$ integration time. The same cases are plotted for $t_\text{Int} = 50\,\mu\text{s}$ on the right. The lowest beam energy shown in both graphs corresponds to an energy {deposited to the detector which is} equivalent to the upper limit of the ePix100a dynamic range {within one integration cycle}, i.e. 100 photons at $8\,\text{keV}$. For the estimate we assumed that photons are impinging mostly the same region of the detector during scientific experiments. This assumption is reasonable for small angle scattering experiments. As an example the horizontal lines in both plots visualize the number of hours the detector can be exposed to the beam during one, three and five years of operation at the European XFEL to reach the corresponding dynamic range reduction. For our estimate we assumed $4216$ hours of beam time operation per calendar year at the European XFEL. This value corresponds to the planned X-ray delivery time for the year $2021$. As one beamline serves two scientific instruments, the allocated time is assumed to be shared equally between the two. Moreover, we estimate the detector to be exposed to X-rays only $50\%$ of the available time{, i.e. the detector will be exposed to radiation for $4216\,\text{h}/\text{year} \times 0.5 \times 0.5 =1054\,\text{h}/\text{year}$. Hence the black line visualizing the dynamic range reduction during one year of operation corresponds to $1054$ hours of beam on the detector}. The numbers presented in figure \ref{fig:LifetimeRange} are of general nature and can be transferred to any other usage scenario, e.g. at other X-ray facilities.
Significant reduction of the dynamic range is not expected, if the detector is illuminated with a beam energy below the dynamic range of the ADC, i.e. $\leq 100 \times 8\,\text{keV}$ photons.

\begin{figure}
\centering
\includegraphics[width=0.49\textwidth]{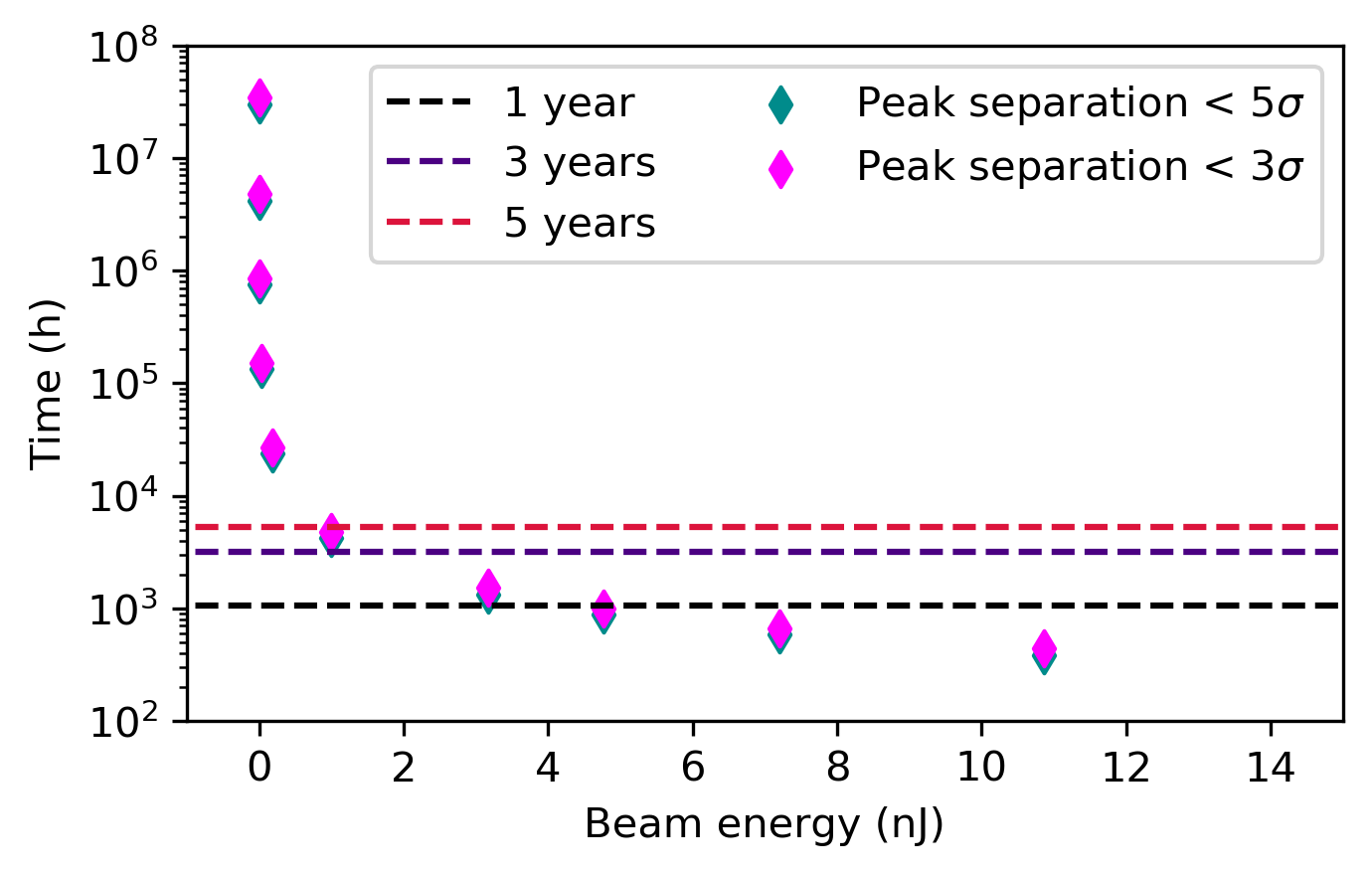}
\hfill
\includegraphics[width=0.49\textwidth]{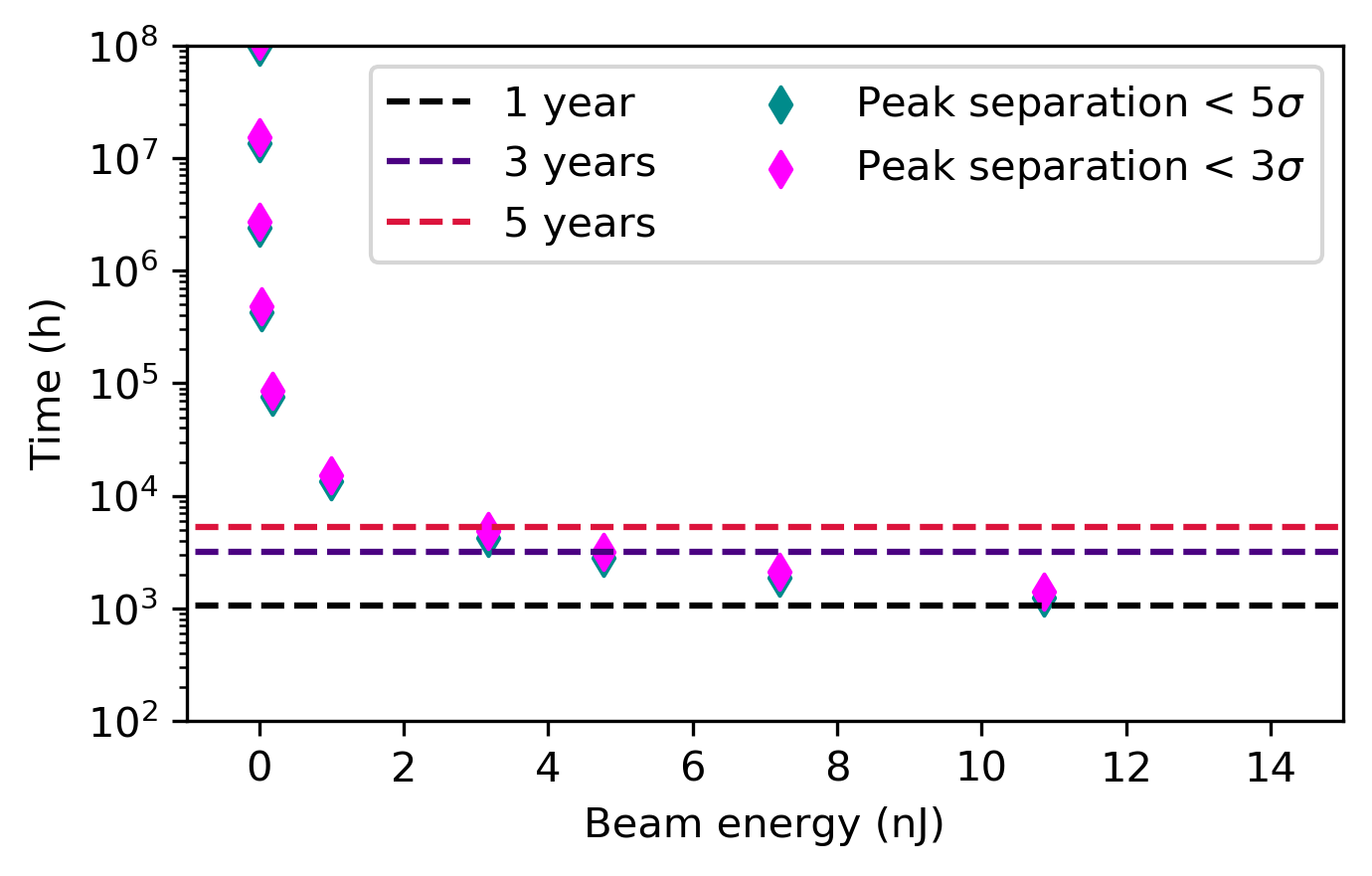}
\caption{Estimate of time needed to reduce a peak-to-noise separation power below $5\,\sigma$ (green) and $3\,\sigma$ (pink) for $t_\text{Int} = 800\,\mu\text{s}$, shown on the left plot and for $t_\text{Int} = 50\,\mu\text{s}$, shown on the right in dependency of the used beam energy.}
\label{fig:LifetimeSeparation}
\end{figure}

The ePix100a detector was designed for low noise spectroscopy applications, hence requiring single photon sensitivity, i.e. good photon-to-noise discrimination down to the lowest photon energies. In this context the detector's noise is an important performance parameter. A common requirement for imaging detectors used at FEL facilities is a false hit detection probability per megapixel area, i.e. $P(0|1)< 10^{-6}$, which corresponds to a photon peak-to-noise separation of approximately $5\,\sigma$ at a given energy. The lowest acceptable signal-to-noise value is usually considered to be $3\,\sigma$.
Figure~\ref{fig:LifetimeSeparation} shows the evolution of the signal to noise ratio if the detector is exposed to a given beam energy for a specific amount of time. The $5\,\sigma$ (cyan) or $3\,\sigma$ (magenta) peak separation are indicated for a photon energy of $9\,\text{keV}$ . The plot on the left shows the separation power reduction at $800\,\mu\text{s}$ integration time and the right plot at the integration time of $50\,\mu\text{s}$. As in the previous figure, the horizontal lines mark the  beam time hours at the European XFEL per calendar year. A critical noise increase, hence reduction in peak-to-noise separation is only expected at beam intensities above the detector's dynamic range. Irradiating the detector for $2\,\text{years}$ with an energy of $2.5\,\text{nJ}$ would cause a drop of the signal-to-noise ratio below $3\,\sigma$ at $800\,\mu\text{s}$ integration time, while at $50\,\mu\text{s}$ the same energy during $5\,\text{years}$ would lead to a drop below $5\,\sigma$. {The dose deposited to the $\text{SiO}_2$ layers delivered to the area of $20\,\text{pixels} \times 20\,\text{pixels}$ resulting from the beam energy of $2.5\,\text{nJ}$ corresponds to $(1.1 \pm 0.1)\,\text{GGy}$. The same beam energy leads to a dose of $(2.7 \pm 0.2)\,\text{GGy}$ in 5 years. The beam energy needed to worsen the peak-to-noise separation below $5\,\sigma$ in 5 years of operation at $800\,\mu\text{s}$ integration time is $\approx 0.4\,\text{nJ}$, which exceeds the dynamic range of the ePix100a by $\approx 3$ orders of magnitude.} 
Table~\ref{tab:LimitDoses} summarizes the dose thresholds deposited at the $\text{SiO}_2$ above which the detector's dynamic range and peak separation power will 
degrade.

\begin{table}
\centering
\caption{Estimated dose thresholds above which a significant degradation of the detectors' performance will occur, i.e. a reduction of its dynamic range and peak separation power.}
\label{tab:LimitDoses}
\smallskip
\begin{tabular}{crrrr} 
\hline\hline
\multicolumn{1}{l}{} & 
\multicolumn{4}{c}{Dose absorbed in $\text{SiO}_2$}\\ 
\cline{2-5}
Integration time & 
\multicolumn{2}{c}{ADC range reduction} &               \multicolumn{2}{c}{Peak separation}\\ 
\hline
\multicolumn{1}{l}{} & \multicolumn{1}{c}{$50\,\%$} & \multicolumn{1}{c}{$100\,\%$} & \multicolumn{1}{c}{$<5\,\sigma$} & \multicolumn{1}{c}{$<3\,\sigma$}\\ 
\cline{2-5}
\multicolumn{1}{r}{
    $50\,\mu\text{s}$}           & 
    $(7.4{\pm 1.0)}\,\text{MGy}$ & $(14.8{\pm 2)}\,\text{MGy}$ & 
$(28{\pm 4)}\,\text{kGy}$ & $(32{\pm 4)}\,\text{kGy}$\\
\multicolumn{1}{r}{$800\,\mu\text{s}$} & $(131{\pm 18)}\,\text{kGy}$ & $(262{\pm 36)}\,\text{kGy}$ & $(9{\pm 1)}\,\text{kGy}$                 
& $(10{\pm 1})\,\text{kGy}$\\ 
\hline\hline
\end{tabular}
\end{table}

\section{Conclusions \& Outlook}
We have performed a systematic study of the influence of radiation induced damage on the performance of the ePix100a detector. We irradiated the ePix100a detector under controlled conditions with the direct and attenuated European XFEL beam with X-ray photons with an energy of $9\,\text{keV}$ and a beam energy of $1\,\mu\text{J}$. Pixels irradiated by this energy do not show a signal dependent response upon irradiation but remain functional under normal operating conditions. {Irradiating the detector} beyond the beam energy of $1\,\mu\text{J}$ {for longer time periods, e.g. $> 5\,\text{min}$} will cause {failure of} the irradiated pixels. Furthermore, we provide irradiation limits for typical usage scenarios at the European XFEL, which results in a certain damage level. Our results can be transferred to experimental conditions at other facilities and experiments.

The irradiated area of $1\,\text{mm}^2$ has received a dose of approximately $760\,\text{kGy}$ at the depth of $\text{Si}/\text{SiO}_2$ in the sensor, which corresponds to $180\,\text{MGy}$ delivered to the surface of the sensor. The dose dependent increase of the offset and noise is mainly caused by an increase of the leakage current. The observed broadening of the {$\text{Cu-K}_{\alpha}$} fluorescence line measured $90\,\text{min}$ post irradiation is scaling with the increasing noise in the pixels and thus is caused by the radiation-induced leakage current. A change of the gain is not expected for a dose $< 4\,\text{kGy}$. Nevertheless, a charge injection scan showed a weak gain decrease for the most irradiated pixels and suggests a weak damage occurring at the pixel preamplifier. Single photon discrimination at a significance level of $>5\,\sigma$ can be achieved with the ePix100a up to a dose of $9\,\text{kGy}$ at $t_\text{Int} = 800\,\mu\text{s}$ and up to $28\,\text{kGy}$ at $t_\text{Int} =50\,\mu\text{s}$. 

In the near future, we plan to investigate sensor annealing as a possibility to mitigate the radiation induced performance changes to conclude the ePix100a radiation hardness study.

\acknowledgments
We acknowledge the European XFEL in Schenefeld, Germany, for provision of a X-ray free-electron laser beamtime at the HED instrument and would like to thank the beam line staff for their assistance.

The work presented in this publication was funded by the European XFEL. We would like to thank specifically the following European XFEL groups for their fruitful collaboration, vital contribution to this work and their continuous effort in supporting this project: Control devices were developed by the Controls group led by Darren Spruce, data acquisition and storage is provided by the Information Technology and Data Management (ITDM) group led by Krzysztof Wrona. We would like to thank Theophilos Maltezopoulos from the X-ray Photon Diagnostics (XPD) group for his support in analyzing the XGM data and Dionisio Doering and Maciej Kwiatkowski for support with execution of the charge injection scan.


\bibliography{epix-radiation}
\bibliographystyle{unsrt}

\end{document}